\documentclass[aps,twocolumn,showpacs,preprintnumbers,amsmath,amssymb,
nofootinbib,showkeys,floatfix]{revtex4-1}

\usepackage[english]{babel}
\usepackage[utf8x]{inputenc}

\usepackage{epsfig}
\usepackage{graphicx}
\usepackage{dcolumn}
\usepackage{latexsym}
\usepackage{graphicx}
\usepackage{enumerate}
\usepackage{float}
\usepackage{placeins}
\usepackage{supertabular}

\usepackage{hyperref} 
\hypersetup{
   colorlinks=true,
   linkcolor=blue,
   citecolor=blue,
   urlcolor=blue
}

\def\tstrut{\vrule height2.5ex depth0pt width0pt} % used in tables

\bibpunct{[}{]}{,}{n}{,}{,} % options of natbib

\begin{document}

\title{Bottomonium spectrum revisited}
\author{Jorge Segovia}
\email{jorge.segovia@tum.de}
\affiliation{Physik-Department, Technische Universit\"at M\"unchen, 
James-Franck-Str. 1, 85748 Garching, Germany}
\affiliation{Instituto Universitario de F\'isica Fundamental y 
Matem\'aticas (IUFFyM) \\ Universidad de Salamanca, E-37008 Salamanca, Spain}
\author{Pablo G. Ortega}
\email{pgarciao@cern.ch}
\affiliation{CERN (European Organization for Nuclear Research), CH-1211
Geneva, Switzerland}
\author{David R. Entem}
\email{entem@usal.es}
\author{Francisco Fern\'andez}
\email{fdz@usal.es}
\affiliation{Grupo de F\'isica Nuclear and Instituto Universitario de
F\'isica Fundamental y Matem\'aticas (IUFFyM) \\ Universidad de Salamanca,
E-37008 Salamanca, Spain}
\date{\today}

\begin{abstract}
We revisit the bottomonium spectrum motivated by the recently exciting 
experimental progress in the observation of new bottomonium states, both 
conventional and unconventional. Our framework is a nonrelativistic constituent 
quark model which has been applied to a wide range of hadronic observables from 
the light to the heavy quark sector and thus the model parameters are 
completely constrained. Beyond the spectrum, we provide a large number of 
electromagnetic, strong and hadronic decays in order to discuss the quark 
content of the bottomonium states and give more insights about the better way 
to determine their properties experimentally.
\end{abstract}


\maketitle

% 3) Incluir al calculo molecular de \Upsilon(4S).
% 4) Actualizar el summary que haremos al final del todo.

\section{INTRODUCTION}
\label{sec:introduction}

\subsection{Experimental situation}
\label{subsec:experiment}

Bottomonium, a bound system of a bottom $(b)$ quark and its antiquark
$(\bar{b})$, was discovered as spin-triplet states called $\Upsilon(1S)$,
$\Upsilon(2S)$ and $\Upsilon(3S)$ by the E288 Collaboration at Fermilab in 1977
in proton scattering on $Cu$ and $Pb$ targets studying muon pairs in a regime of
invariant masses larger than $5\,{\rm GeV}$~\cite{Herb:1977ek,Innes:1977ae}.
Later, they were better studied at various $e^{+}e^{-}$ storage rings. The two
triplet $P$-wave states $\chi_{bJ}(2P)$ and $\chi_{bJ}(1P)$ with $J=0,\,1,\,2$
were discovered in radiative decays of the $\Upsilon(3S)$ and $\Upsilon(2S)$ in
$1982$~\cite{Han:1982zk,Eigen:1982zm} and
$1983$~\cite{Klopfenstein:1983nx,Pauss:1983pa}, respectively.

Despite such early measurements, during the next thirty years there were no
significant contributions to the spectrum of bottomonium. Only the radial
excitations of the vector bottomonium family $\Upsilon(4S)$, $\Upsilon(10860)$,
and $\Upsilon(11020)$ were observed~\cite{Besson:1984bd,Lovelock:1985nb}. This
was largely because the $B$-factories were not usually considered ideal
facilities for the study of the bottomonium spectrum since their energy was
tuned to the peak of the $\Upsilon(4S)$ resonance, which decays in almost
$100\%$ of cases to a $B\bar{B}$ pair.

The situation has changed dramatically in the last few years with many
bottomonium states observed. In $2008$, the spin-singlet pseudoscalar partner
$\eta_{b}(1S)$ was found by the BaBar Collaboration with a mass of
$9388.9^{+3.1}_{-2.3}\pm2.7\,{\rm MeV}$~\cite{Aubert:2008ba}. A second
measurement of BaBar found a figure slightly higher $9394^{+4.8}_{-4.9}\,{\rm
MeV}$ but perfectly compatible. A later measurement of the
CLEO~\cite{Bonvicini:2009hs} Collaboration gave a value of $9391\pm6.6\,{\rm
MeV}$. The Belle Collaboration~\cite{Mizuk:2012pb}, with a simultaneous fit of 
its mass and width, obtains a value of 
$m_{\eta_{b}(1S)}=(9402.4\pm1.5\pm1.8)\,{\rm MeV}$ for the
mass and $\Gamma_{\eta_{b}(1S)}=(10.8^{+4.0+4.5}_{-3.7-2.0})\,{\rm MeV}$ for the
width. This is the most precise measurement of the $\eta_{b}(1S)$ mass and
furthermore provides its total decay width.

In Ref.~\cite{Lees:2011mx} the BaBar Collaboration searched for radiative 
decays to the $\eta_{b}(1S)$ and $\eta_{b}(2S)$ states. Despite of their 
results are largely inconclusive, they observed a signal of the $\eta_{b}(2S)$ 
state with a mass over a range of approximately 
$9974<m_{\eta_{b}(2S)}<10015\,{\rm MeV}$. Then, the CLEO Collaboration 
presented evidence for the first successful observation of $\eta_{b}(2S)$ in 
$\Upsilon(2S)\to\eta_{b}(2S)\gamma$ decays at a mass of 
$9974.6\pm2.3\pm2.1\,{\rm MeV}$~\cite{Dobbs:2012zn}. And soon after that, the 
Belle Collaboration~\cite{Mizuk:2012pb} reported a signal for the 
$\eta_{b}(2S)$ using the $h_{b}(2P)\to \eta_{b}(2S)\gamma$ transition at a mass 
of $(9999.0\pm3.5^{+2.8}_{-1.9})\,{\rm MeV}$. This value is clearly incompatible
with the previous one. An analysis performed by Belle~\cite{Sandilya:2013rhy} 
with almost $17$ times more data found no evidence for a signal in the energy
region around $9975\,{\rm MeV}$, casting doubt on the CLEO result.

Experimentalists have been only able to distinguish the $\Upsilon(1^{3}D_{2})$
state of the triplet 
$\Upsilon(1^{3}D_{J})$~\cite{Bonvicini:2004yj,delAmoSanchez:2010kz}. In
Ref.~\cite{delAmoSanchez:2010kz} the $J=2$ member of the $\Upsilon(1^{3}D_{J})$
spin-triplet was observed through the $\Upsilon(3S)\to
\gamma\gamma\Upsilon(1^{3}D_{J})\to \gamma\gamma\pi^{+}\pi^{-}\Upsilon(1S)$
decay chain with a significance of $5.8$ standard deviations including
systematic uncertainties. For the other two members of this spin-triplet,
$\Upsilon(1^{3}D_{1})$ and $\Upsilon(1^{3}D_{3})$, the significances were much
lower, $1.8$ and $1.6$ respectively, and thus no experimental observation can be
claimed.

Evidence for the lowest spin-singlet $P$-wave state, $h_{b}(1P)$, was first
reported by the BaBar Collaboration in the transition $\Upsilon(3S)\to
\pi^{0}h_{b}(1P)\to \pi^{0}\gamma\eta_{b}(1S)$~\cite{Lees:2011zp}. They found a
$3\sigma$ excess of events in the recoil mass distribution against $\pi^{0}$ at
a mass of $(9902\pm4\pm2)\,{\rm MeV}$. This spin singlet $P$-wave state is 
expected to be very close in mass to the spin weighted average of the triplet 
states $\left\langle m(1^{3}P_{J})\right\rangle = 9899.9\,{\rm MeV}$. The first
significant signal for this state come from the Belle Collaboration in the
$\Upsilon(5S)\to h_{b}(1P)\pi^{+}\pi^{-}$ transition~\cite{Adachi:2011ji}. They
were also able to distinguish its first radial excitation, $h_{b}(2P)$. The
measured masses of the $h_{b}(1P)$ and $h_{b}(2P)$ states were
$9898.25\pm1.06^{+1.03}_{-1.07}\,{\rm MeV}$ and
$10259.76\pm0.64^{+1.43}_{-1.03}\,{\rm MeV}$.

The proton--(anti-)proton colliders have joined recently in the search of 
bottomonium states. A clear example is the observation of the $\chi_{bJ}(nP)$ 
states produced in proton-proton collisions at the LHC at $\sqrt{s}=7\,{\rm 
TeV}$ and recorded by the ATLAS detector~\cite{Aad:2011ih}. These states have 
been reconstructed through their radiative decays to $\Upsilon(1S,2S)$ with
$\Upsilon\to \mu^{+}\mu^{-}$. In addition to the mass peaks corresponding to the
decay modes $\chi_{bJ}(1P,2P)\to \Upsilon(1S)\gamma$, a new structure centered
at a mass of $10.530\pm0.005\pm0.009\,{\rm GeV}$ has been also observed, in both
the $\Upsilon(1S)\gamma$ and $\Upsilon(2S)\gamma$ decay modes. This structure
has been assigned to the $\chi_{bJ}(3P)$ system. Soon after that, the D0 
Collaboration observed a peak in the $\Upsilon(1S)\gamma$ final state at a mass
of $10.551\pm0.014\pm0.017\,{\rm GeV}$~\cite{aba:2012} which is compatible with
the new state observed by the ATLAS Collaboration. The LHCb Collaboration has 
recently determined the mass of the $\chi_{b1}(3P)$ to be 
$m(\chi_{b1}(3P))=10515^{+2.2}_{-3.9}{\rm 
(stat)}^{+1.5}_{-2.1}{\rm(syst)}\,{\rm MeV}$~\cite{Aaij:2014hla}.

We have discussed in this section the whole spectrum of bottomonium reported
in $2014$ by the Particle Data Group (PDG)~\cite{PDG2014} except two states: the
$X(10610)^{\pm}$ and $X(10650)^{\pm}$. These two states were observed by the
Belle Collaboration~\cite{Belle:2011aa} in the mass spectra of the
$\pi^{\pm}\Upsilon(nS)$ $(n=1,\,2,\,3)$ and $\pi^{\pm}h_{b}(mP)$ $(m=1,\,2)$
pairs that are produced in association with a single charged pion in
$\Upsilon(5S)$ decays. The measured masses and widths of the two structures
averaged over the five final states are $(10607.2\pm2.0)\,{\rm MeV}$ and
$(18.4\pm2.4)\,{\rm MeV}$ for the $X(10610)^{\pm}$, and $(10652.2\pm1.5)\,{\rm
MeV}$ and $(11.5\pm2.2)\,{\rm MeV}$ for $X(10650)^{\pm}$. Their large mass,
indicating the presence of two bottom quarks in their composition, together with
their possession of electrical charge, marks these states as necessarily
unconventional. Their proximity to the $BB^{\ast}$ and $B^{\ast}B^{\ast}$
thresholds makes their identification as molecular states an attractive
possibility. However, there is a strong discussion within the scientific
community about other possible interpretations~\cite{Bugg:2011jr, 
Danilkin:2011sh, Cui:2011fj, Guo:2011gu}.

%%%%%%%%%%%%%%%%%%%%%%%%%%%%%%%%%%%%%%%%%%%%%%%%%%%%%%%%%%%%%%%%%%%%%%%%%%%%%%%%
%%%%%%%%%%%%%%%%%%%%%%%%%%%%%%%%%%%%%%%%%%%%%%%%%%%%%%%%%%%%%%%%%%%%%%%%%%%%%%%%

\subsection{Theoretical tool(s)}
\label{subsec:theory}

The description of hadrons containing two heavy quarks is a rather challenging
problem from the point of view of QCD. One has to add the complications
of a nonperturbative low-energy dynamics to those usually coming from solving
the bound state problem in quantum field theory. 

A proper relativistic quantum field theoretical treatment of the heavy
quarkonium system based on the Bethe-Salpeter equation has proved to be
difficult despite its relatively success in the last few 
years~\cite{Gomez-Rocha:2014vsa, Hilger:2014nma, Fischer:2014cfa}.

The two most promising approaches to the bottomonium bound state problem are
Effective Field Theories (EFTs) and lattice gauge theories. EFTs directly
derived from QCD, like nonrelativistic QCD (NRQCD)~\cite{Caswell:1985ui, 
Bodwin:1994jh} or potential nonrelativistic QCD (pNRQCD)~\cite{Pineda:1997bj, 
Brambilla:1999xf} (for some reviews see Refs.~\cite{Brambilla:2004jw, 
Pineda:2011dg}), disentangle the dynamics of the heavy quarks from the dynamics 
of the light degrees of freedom efficiently and in a model-independent way. The 
fully relativistic dynamics can, in principle, be treated without approximations 
in lattice gauge theories (see, for instance, Ref.~\cite{Dudek:2007wv} for a 
standard lattice heavy quarkonium spectrum). Heavy quark calculations within 
these two approaches have experienced a considerable progress for states away 
from threshold. However, the threshold regions remain troublesome for the EFTs 
as well as lattice-regularized QCD~\cite{Brambilla:2010cs}. Moreover, the 
lattice calculations of excited states have been only recently pioneered and the 
full treatment of bottomonium on the lattice seems to be tricky (for a global 
picture on lattice-regularized QCD calculations and their complexities in the 
bottomonium sector, the reader is referred to~\cite{Gray:2005ur, Meinel:2009rd, 
Burch:2009az, Meinel:2010pv, Dowdall:2011wh, Lewis:2012ir} and references 
therein).
 
All this together explains why many of our expectations in heavy quarkonia still
rely on potential models. Potential formulations have been successful at
describing the heavy quark-antiquark system since the early days of charmonium 
(see {\it e.g.}~\cite{Eichten:1978tg, Eichten:1979ms, Gupta:1982kp, 
Gupta:1983we, Gupta:1984jb, Gupta:1984um, Kwong:1987ak, Kwong:1988ae, 
Barnes:1996ff, Ebert:2002pp, Eichten:2007qx, Danilkin:2009hr, Ferretti:2013vua, 
Godfrey:2015dia}). Moreover, the predictions within this formalism of heavy 
quarkonium properties related with decays and reactions have turned to be very 
valuable for experimental searches. One can mention, for instance, the 
remarkable success of the $^{3}P_{0}$ strong decay model~\cite{Micu:1968mk, 
LeYaouanc:1972ae, LeYaouanc:1973xz, Capstick:1986bm, Capstick:1993kb, 
Page:1995rh, Ackleh:1996yt}. Finally, the easy way to extend the quark model 
for describing multiquark systems makes this framework a suitable one for 
exploratory purposes. The results presented herein are based on a derived 
version of this approach: a nonrelativistic constituent quark model (CQM).

Spontaneous chiral symmetry breaking of the QCD Lagrangian together with the
perturbative one-gluon exchange (OGE) and the nonperturbative confining
interaction are the main pieces of constituent quark models. Using this idea,
Vijande {\it et al.}~\cite{Vijande:2004he} developed a model of the 
quark-antiquark interaction which is able to describe meson phenomenology from 
the light to the heavy quark sector. We have adopted this model and fine tune 
its parameters to reproduce the highest excited states that appear in the light 
quark sector~\cite{Segovia:2008zza}. The reason for that lies in the fact that 
it is widely believed that confinement is flavor independent. Therefore, the
interactions which largely determine the high energy spectrum of heavy quarkonia
should be constrained also by the light quark sector.

The quark model parameters relevant for this work are shown in 
Table~\ref{tab:parameters}. In the heavy quark sector chiral symmetry is 
explicitly broken and Goldstone-boson exchanges do not appear. Thus, OGE and 
confinement are the only interactions remaining. Explicit expressions of these 
interactions and a brief description of the potential is given in 
Appendix~\ref{app:CQM}. Further details about the quark model and the 
fine-tuned model parameters can be found in 
Refs.~\cite{Vijande:2004he,Segovia:2008zza,Segovia:2008zz}. 

\begin{table}[!t]
\begin{center}
\begin{tabular}{lcc}
\hline
\hline
\tstrut
Quark masses & $m_{n}$ (MeV) & $313$ \\
             & $m_{s}$ (MeV) & $555$ \\
             & $m_{b}$ (MeV) & $5110$ \\
\hline
\tstrut
OGE & $\hat{r}_{0}$ (fm) & $0.181$ \\
    & $\hat{r}_{g}$ (fm) & $0.259$ \\
    & $\alpha_{0}$ & $2.118$ \\
    & $\Lambda_{0}$ $(\mbox{fm}^{-1})$ & $0.113$ \\
    & $\mu_{0}$ (MeV) & $36.976$ \\
\hline
\tstrut
Confinement & $a_{c}$ (MeV) & $507.4$ \\
            & $\mu_{c}$ $(\mbox{fm}^{-1})$ & $0.576$ \\
            & $\Delta$ (MeV) & $184.432$ \\
            & $a_{s}$ & $0.81$ \\
\hline
\hline
\end{tabular}
\caption{\label{tab:parameters} Quark model parameters.}
\end{center}
\end{table}

Masses of meson states are a relevant piece of information about their
structure. However, a more complete description can be achieved studying mesonic
decays. In this way, we are checking particular regions of the wave function and
not an average over the all meson size as in the calculation of the mass
spectrum. Appendix~\ref{app:decays} provides the necessary formulation to carry
out the calculations presented herein on annihilation processes and on
electromagnetic, strong and hadronic decays.

Two sections of this manuscript are still to be introduced. In 
Sec.~\ref{sec:results} we discuss our quark model results and compare them with 
the available experimental data. We finish summarizing and giving some 
conclusions in Sec.~\ref{sec:conclusions}.

%%%%%%%%%%%%%%%%%%%%%%%%%%%%%%%%%%%%%%%%%%%%%%%%%%%%%%%%%%%%%%%%%%%%%%%%%%%%%%%%
%%%%%%%%%%%%%%%%%%%%%%%%%%%%%%%%%%%%%%%%%%%%%%%%%%%%%%%%%%%%%%%%%%%%%%%%%%%%%%%%

\begin{table}[!t]
\begin{center}
\scalebox{0.99}{\begin{tabular}{cccccc}
\hline
\hline
\tstrut
State & $J^{PC}$ & $nL$ & The. (MeV) & Exp. (MeV) & \cite{PDG2014} \\
\hline
\tstrut
$\eta_{b}$     & $0^{-+}$ & $1S$ & $9455$  & $9398.0\pm3.2$ & \\
               &          & $2S$ & $9990$  & $9999.0\pm3.5^{+2.8}_{-1.9}$ & \\
               &          & $3S$ & $10330$ & - & \\[2ex]
$\chi_{b0}$    & $0^{++}$ & $1P$ & $9855$  & $9859.44\pm0.42\pm0.31$ & \\
               &          & $2P$ & $10221$ & $10232.5\pm0.4\pm0.5$ & \\
               &          & $3P$ & $10500$ & - & \\[2ex]
$h_{b}$        & $1^{+-}$ & $1P$ & $9879$  & $9899.3\pm1.0$ & \\
               &          & $2P$ & $10240$ & $10259.8\pm0.5\pm1.1$ & \\
               &          & $3P$ & $10516$ & - & \\[2ex]
$\Upsilon$     & $1^{--}$ & $1S$ & $9502$  & $9460.30\pm0.26$ & \\
               &          & $2S$ & $10015$ & $10023.26\pm0.31$ & \\
               &          & $1D$ & $10117$ & - & \\
               &          & $3S$ & $10349$ & $10355.2\pm0.5$ & \\ 
               &          & $2D$ & $10414$ & - & \\
               &          & $4S$ & $10607$ & $10579.4\pm1.2$ & \\
               &          & $3D$ & $10653$ & - & \\
               &          & $5S$ & $10818$ & $10876\pm11$ & \\
               &          & $4D$ & $10853$ & - & \\
               &          & $6S$ & $10995$ & $11019\pm8$ & \\
               &          & $5D$ & $11023$ & - & \\[2ex]
$\chi_{b1}$    & $1^{++}$ & $1P$ & $9874$  & $9892.78\pm0.26\pm0.31$ & \\
               &          & $2P$ & $10236$ & $10255.46\pm0.22\pm0.50$ & \\
               &          & $3P$ & $10513$ & $10515.7^{+2.2+1.5}_{-3.9-2.1}$ & 
\cite{Aaij:2014hla} \\[2ex]
$\eta_{b2}$    & $2^{-+}$ & $1D$ & $10123$ & - & \\
               &          & $2D$ & $10419$ & - & \\
               &          & $3D$ & $10658$ & - & \\[2ex]
$\chi_{b2}$    & $2^{++}$ & $1P$ & $9886$  & $9912.21\pm0.26\pm0.31$ & \\
               &          & $2P$ & $10246$ & $10268.65\pm0.22\pm0.50$ & \\
               &          & $1F$ & $10315$ & - & \\
               &          & $3P$ & $10521$ & - & \\
               &          & $2F$ & $10569$ & - & \\
               &          & $4P$ & $10744$ & - & \\
               &          & $3F$ & $10782$ & - & \\[2ex]
$\Upsilon_{2}$ & $2^{--}$ & $1D$ & $10122$ & $10163.7\pm1.4$ & \\
               &          & $2D$ & $10418$ & - & \\
               &          & $3D$ & $10657$ & - & \\[2ex]
$h_{b3}$       & $3^{+-}$ & $1F$ & $10322$ & - & \\
               &          & $2F$ & $10573$ & - & \\
               &          & $3F$ & $10785$ & - & \\[2ex]
$\Upsilon_{3}$ & $3^{--}$ & $1D$ & $10127$ & - & \\
               &          & $2D$ & $10422$ & - & \\
               &          & $1G$ & $10506$ & - & \\
               &          & $3D$ & $10660$ & - & \\
               &          & $2G$ & $10712$ & - & \\
               &          & $4D$ & $10860$ & - & \\
               &          & $3G$ & $10904$ & - & \\[2ex]
$\chi_{b3}$    & $3^{++}$ & $1F$ & $10321$ & - & \\
               &          & $2F$ & $10573$ & - & \\
               &          & $3F$ & $10785$ & - & \\
\hline
\hline
\end{tabular}}
\caption{\label{tab:predmassesbb} Masses, in MeV, of bottomonium states (up to 
spin $J=3$) predicted by our constituent quark model.}
\end{center}
\end{table}

\section{RESULTS AND DISCUSSION}
\label{sec:results}

Table~\ref{tab:predmassesbb} shows the bottomonium spectrum (up to spin $J=3$)
predicted by our constituent quark model. All world average masses reported in
the PDG~\cite{PDG2014} are also shown. The masses for those states that are 
not yet considered as well established by PDG have been taken from the original
experimental works. It is inferred from Table~\ref{tab:predmassesbb} that a 
global description of the bottomonium spectrum is obtained by our CQM. A 
detailed discussion about the particular features of our spectrum will be given 
in the following subsections. We shall also compute decay properties of the 
studied mesons.

\subsection{The $\mathbf{\eta_{b}}$ states}

Table~\ref{tab:predmassesbb} shows the predicted masses of the $\eta_{b}$ 
states. The hyperfine mass-splitting of singlet-triplet states, {\it i.e.} 
${\Delta m}_{hf}[\eta_{b}(nS)] = m(n^{3}S_{1})-m(n^{1}S_{0})$, probes the 
spin-dependence of bound-state energy levels and imposes constraints on 
theoretical descriptions. For the ground states, $n=1$, it is given 
experimentally by~\cite{PDG2014}
\begin{equation}
{\Delta m}_{hf}[\eta_{b}(1S)]=62.3\pm3.2\,{\rm MeV},
\end{equation}
which is higher than the theoretical prediction of EFTs, 
$41\pm11^{+9}_{-8}$~\cite{Kniehl:2003ap}, and compatible with the lattice 
regularized QCD result, $(60.3\pm7.7)\,{\rm MeV}$~\cite{Meinel:2010pv}. The 
Belle Collaboration, with a simultaneous fit of the mass and width of the 
$\eta_{b}(1S)$, reduces this hyperfine splitting and obtains a value of 
$(57.9\pm2.3)\,{\rm MeV}$~\cite{Mizuk:2012pb}. The hyperfine mass splitting 
predicted by our quark model is $47\,{\rm MeV}$, higher than the result obtained 
by EFTs but still lower than the experimental data and lattice regularized QCD 
computation.

We predict for the $\eta_{b}(2S)$ state a mass of $9990\,{\rm MeV}$ which is in 
good agreement with the last experimental measurement performed by the Belle 
Collaboration~\cite{Mizuk:2012pb}, $(9999.0\pm3.5^{+2.8}_{-1.9})\,{\rm MeV}$. 
The corresponding theoretical hyperfine mass splitting ${\Delta 
m}_{hf}[\eta_{b}(2S)]=25\,{\rm MeV}$ is in excellent agreement with the 
experimental one ${\Delta m}_{hf}[\eta_{b}(2S)]=24.3^{+4.0}_{-4.5}\,{\rm 
MeV}$~\cite{Mizuk:2012pb}. It is worth to mention that our splitting is also in 
very good agreement with the latest results of lattice regularized QCD, ${\Delta 
m}_{hf}[\eta_{b}(2S)]=(23.5-28.0)\,{\rm MeV}$~\cite{Meinel:2010pv}.

One can see in Table~\ref{tab:predmassesbb} that our predicted mass for the 
$\eta_{b}(3S)$ is $10330\,{\rm MeV}$. Its corresponding hyperfine mass 
splitting is ${\Delta m}_{hf}[\eta_{b}(3S)]=19\,{\rm MeV}$, which follows the
expected trend. 

The decay widths and branching fractions of annihilation rates, radiative 
decays and hadronic transitions for the $\eta_{b}$ states are given in 
Table~\ref{tab:E1etab}. In all cases, the most important contribution to the 
total decay width comes from the gluon annihilation rate. Note that they are 
given in MeV whereas the rest of the quantities are given in keV. Beyond the 
gluon annihilation rates, the next significant decay channel shown in 
Table~\ref{tab:E1etab} is the $\eta_{b}(2S)$ into $\pi\pi\eta_{b}(1S)$ with a 
width four times higher than the most dominant electromagnetic transitions 
$\eta_{b}(2S)\to h_{b}(1P)\gamma$ and $\eta_{b}(3S)\to h_{b}(2P)\gamma$. This 
behavior is due to the fact that we use the full expression of 
Eq.~(\ref{eq:EfiE1}) for the E1 radiative decays and not only the leading 
term. The corrections over the low energy expansion are important for these 
excited states. The hadronic transition of the $\eta_{b}(3S)$ into the 
$\pi\pi\eta_{b}(1S)$ final channel presents a decay width of the same order 
of magnitude than that of the $\eta_{b}(3S)\to h_{b}(2P)\gamma$ decay.

It is important to mention here that, as we will explain next, we do not expect 
perfect agreement with experiment for the pseudoscalar mesons. The worst 
situation is found for the $\eta_{b}(1S)$ state and then it is alleviated with 
higher excitations. This trend is inferred from the discussion above, and the 
most evidence of this appears in our calculation of the total decay width of 
the $\eta_{b}(1S)$ state, $20.2\,{\rm MeV}$, which is a factor of two larger 
than the central value of the last experimental 
measurement~\cite{Mizuk:2012pb}, $(10.8^{+4.0+4.5}_{-3.7-2.0})\,{\rm MeV}$ (see 
that our theoretical result lies just above the upper limit of the error bar). 
The reason for that is the following: Our CQM presents an OGE potential which 
has a spin-spin contact hyperfine interaction that is proportional to a Dirac 
delta function, conveniently regularized, at the origin (see 
Eqs.~(\ref{eq:OGEpot}) and~(\ref{eq:delta})). The corresponding regularization 
parameter was fitted to determine the hyperfine splittings between the 
$n^{1}S_{0}$ and $n^{3}S_{1}$ states in the different flavor sectors. While 
most of the physical observables are insensitive to the regularization of this 
delta term, those related with annihilation processes are affected because 
these processes are driven by short range operators~\cite{Segovia:2011tb, 
Segovia:2012cs}. The effect is very small in the $^{3}S_{1}$ channel as the 
delta term is repulsive in this case. It is negligible for higher partial waves 
due to the shielding by the centrifugal barrier. However, it is sizable in the 
$^{1}S_{0}$ channel for which the delta term is attractive and the sensitivity 
decreases as going up in higher excited states.

\begin{table}[!t]
\begin{center}
\begin{tabular}{llrr}
\hline
\hline
\tstrut
Initial state & Final state & $\Gamma_{\rm The.}$ & ${\cal B}_{\rm The.}$ \\
& & (keV) & $(10^{-2})$ \\
\hline
\tstrut
$\eta_{b}(1S)$ & $gg$           & $20.18\,{\rm MeV}$ & $\sim100.00$ \\
               & $\gamma\gamma$ & $0.69$             & $3.42\times10^{-3}$ \\
               & total          & $20.18\,{\rm MeV}$ & $100.00$ \\[2ex]
$\eta_{b}(2S)$ & $gg$                 & $10.64\,{\rm MeV}$  & $99.86$ \\
               & $\gamma\gamma$       & $0.36$              & $3.38\times10^{-3}$ \\
               & $\gamma h_{b}(1P)$   & $2.85$              & $2.68\times10^{-2}$ \\
               & $\gamma\Upsilon(1S)$ & $4.50\times10^{-2}$ & $4.22\times10^{-4}$ \\
               & $\pi\pi\eta_{b}(1S)$ & $11.27$             & $10.58\times10^{-2}$ \\
               & total                & $10.66\,{\rm MeV}$  & $100.00$ \\[2ex]
$\eta_{b}(3S)$ & $gg$                 & $7.94\,{\rm MeV}$   & $99.93$ \\
               & $\gamma\gamma$       & $0.27$              & $3.40\times10^{-3}$ \\
               & $\gamma h_{b}(1P)$   & $8.40\times10^{-3}$ & $1.06\times10^{-4}$ \\
               & $\gamma h_{b}(2P)$   & $2.60$              & $3.27\times10^{-2}$ \\
               & $\gamma\Upsilon(1S)$ & $5.10\times10^{-2}$ & $6.42\times10^{-4}$ \\
               & $\gamma\Upsilon(2S)$ & $9.20\times10^{-3}$ & $1.16\times10^{-4}$ \\     
               & $\pi\pi\eta_{b}(1S)$ & $1.95$              & $2.45\times10^{-2}$ \\
               & $\pi\pi\eta_{b}(2S)$ & $0.34$              & $4.28\times10^{-3}$ \\
               & total                & $7.95\,{\rm MeV}$   & $100.00$ \\        
\hline
\hline
\end{tabular}
\caption{\label{tab:E1etab} Decay widths and branching fractions of annihilation 
rates, radiative decays and hadronic transitions for the $\eta_{b}$ states. 
There is no experimental data available.}
\end{center}
\end{table}

%%%%%%%%%%%%%%%%%%%%%%%%%%%%%%%%%%%%%%%%%%%%%%%%%%%%%%%%%%%%%%%%%%%%%%%%%%%%%%%%

\begin{table}[!t]
\begin{center}
\scalebox{0.92}{\begin{tabular}{ccccc}
\hline
\hline
\tstrut
$n$ & $m_{\rm The.}(h_{b})$ & $m_{\rm Exp.}(h_{b})$ \cite{PDG2014} &
$\left\langle\right.\!\! m(n^{3}P_{J})\!\!\left.\right\rangle_{\rm The.}$ &
$\left\langle\right.\!\! m(n^{3}P_{J})\!\!\left.\right\rangle_{\rm Exp.}$ 
\cite{PDG2014} \\
& (MeV) & (MeV) & (MeV) \\
\hline
\tstrut
$1$ & $9879$  & $9899.3\pm1.0$  & $9879$  & $9899.87\pm0.27$ \\
$2$ & $10240$ & $10259.8\pm1.2$ & $10240$ & $10260.20\pm0.36$ \\
$3$ & $10516$ & -               & $10516$ & $10534\pm9$ \\
\hline
\hline
\end{tabular}}
\caption{\label{tab:centroidbb} The theoretical masses, in MeV, of
the ground state and the first two excitations of $h_{b}$, compared with the
spin-averaged centroid, in MeV, of the corresponding triplet $P$-wave states. We
compare with the experimental data collected in PDG~\cite{PDG2014}.}
\end{center}
\end{table}

\subsection{The $\mathbf{h_{b}}$ and $\mathbf{\chi_{bJ}}$ states}

Table~\ref{tab:predmassesbb} shows the predicted masses of the singlet 
$^{1}P_{1}$ and the triplet $^{3}P_{J}$ states. They are in reasonable agreement
with the experimental data.

The spin-singlet $P$-wave states, $h_{b}$, are expected to lie very close in
mass to the spin-weighted average of the triplet $P$-wave states, $\chi_{bJ}$. 
This is because the hyperfine splitting in leading nonrelativistic order is
proportional to the square of the wave function at the origin, which vanishes
for $P$-wave states. In Table~\ref{tab:centroidbb} we compare the centroid of 
$^{3}P_{J}$ states and the corresponding $h_{b}$ mass for the ground state and 
the first two excitations. One can see, on one hand, that the experimental data 
follow the theoretical expectations and, on the other hand, that our spin-spin 
interaction is negligible for $P$-wave states, as should be. 

It is also important to remark that our spin-averaged centroid of the 
$\chi_{bJ}(3P)$ states, $10516\,{\rm MeV}$, is compatible but slightly higher 
than the value collected by PDG~\cite{PDG2014}. However, the centroid is 
expected to be close to the $\chi_{b1}(3P)$ element of the spin-multiplet and 
the experimental measurement of the mass of this state has been recently 
reported by the LHCb~\cite{Aaij:2014hla} Collaboration with a value of 
$10515.7^{+2.2+1.5}_{-3.9-2.1}\,{\rm MeV}$, which is in perfect agreement with 
our prediction. In addition, we calculate the intra-multiplet splittings as
$m_{\chi_{b2}(3P)}-m_{\chi_{b1}(3P)}=8\,{\rm MeV}$ and
$m_{\chi_{b1}(3P)}-m_{\chi_{b0}(3P)}=13\,{\rm MeV}$.

\begin{table}[!t]
\begin{center}
\scalebox{0.95}{\begin{tabular}{llrrc}
\hline
\hline
\tstrut
Initial state & Final state & $\Gamma_{\rm The.}$ & ${\cal B}_{\rm The.}$
& ${\cal B}_{\rm Exp.}$~\cite{Mizuk:2012pb} \\
& & (keV) &  $(\times10^{-2})$ & $(\times10^{-2})$ \\
\hline
\tstrut
$h_{b}(1P)$ & $ggg$                 & $35.26$             & $44.68$              & - \\
            & $\gamma\eta_{b}(1S)$  & $43.66$             & $55.32$              & $49.2\pm5.7^{+5.6}_{-3.3}$ \\
            & $\gamma\chi_{b0}(1P)$ & $8.61\times10^{-4}$ & $1.09\times10^{-3}$  & - \\
            & $\gamma\chi_{b1}(1P)$ & $1.15\times10^{-5}$ & $1.46\times10^{-5}$  & - \\
            & total                 & $78.92$             & $100.00$             & - \\[2ex]
$h_{b}(2P)$ & $ggg$                 & $52.70$             & $57.82$             & - \\
            & $\gamma\eta_{b}(1S)$  & $14.90$             & $16.35$             & $22.3\pm3.8^{+3.1}_{-3.3}$ \\
            & $\gamma\eta_{b}(2S)$  & $17.60$             & $19.31$             & $47.5\pm10.5^{+6.8}_{-7.7}$ \\
            & $\gamma\eta_{b2}(1D)$ & $5.36$              & $5.88$              & - \\
            & $\gamma\chi_{b0}(1P)$ & $3.64\times10^{-2}$ & $3.99\times10^{-2}$ & - \\
            & $\gamma\chi_{b1}(1P)$ & $1.28\times10^{-3}$ & $1.41\times10^{-3}$ & - \\
            & $\gamma\chi_{b2}(1P)$ & $6.91\times10^{-6}$ & $7.58\times10^{-6}$ & - \\
            & $\pi\pi h_{b}(1P)$    & $0.54$              & $0.59$              & - \\
            & total                 & $91.14$             & $100.00$            & - \\[2ex]
$h_{b}(3P)$ & $ggg$                 & $62.16$             & $64.91$             & - \\
            & $\gamma\eta_{b}(1S)$  & $7.96$              & $8.31$              & - \\
            & $\gamma\eta_{b}(2S)$  & $6.86$              & $7.16$              & - \\
            & $\gamma\eta_{b}(3S)$  & $12.27$             & $12.81$             & - \\
            & $\gamma\eta_{b2}(1D)$ & $0.35$              & $0.37$              & - \\
            & $\gamma\eta_{b2}(2D)$ & $4.72$              & $4.93$              & - \\
            & $\gamma\chi_{b0}(1P)$ & $3.77\times10^{-3}$ & $3.94\times10^{-3}$ & - \\
            & $\gamma\chi_{b1}(1P)$ & $1.23\times10^{-3}$ & $1.28\times10^{-3}$ & - \\
            & $\gamma\chi_{b2}(1P)$ & $5.10\times10^{-5}$ & $5.33\times10^{-5}$ & - \\
            & $\gamma\chi_{b0}(2P)$ & $1.71\times10^{-3}$ & $1.79\times10^{-3}$ & - \\
            & $\gamma\chi_{b1}(2P)$ & $5.97\times10^{-4}$ & $6.23\times10^{-4}$ & - \\
            & $\gamma\chi_{b2}(2P)$ & $7.37\times10^{-6}$ & $7.70\times10^{-6}$ & - \\
            & $\pi\pi h_{b}(1P)$    & $1.44$              & $1.50$              & - \\
            & total                 & $95.76$             & $100.00$            & - \\
\hline
\hline
\end{tabular}}
\caption{\label{tab:E1hb} Decay widths and branching fractions of annihilation 
rates, radiative decays and hadronic transitions for the $h_{b}$ states.
The experimental data are from Ref.~\cite{Mizuk:2012pb}.}
\end{center}
\end{table}

The decay widths and branching fractions of annihilation rates, radiative 
decays and hadronic transitions for the $h_{b}$ states are shown in 
Table~\ref{tab:E1hb}. We predict total decay widths of about $100\,{\rm keV}$ 
for the three $h_{b}$ states. Their total decay widths are dominated by their 
annihilation into gluons.

The Belle Collaboration has studied very recently the processes $e^{+}e^{-}\to
\Upsilon(5S)\to h_{b}(nP)\pi^{+}\pi^{-}\to [\eta_{b}(mS)\gamma]\pi^{+}\pi^{-}$
providing branching fractions for the decays $h_{b}(1P)\to \eta_{b}(1S)\gamma$,
$h_{b}(2P)\to \eta_{b}(1S)\gamma$ and $h_{b}(2P)\to
\eta_{b}(2S)\gamma$~\cite{Mizuk:2012pb}. One can see in Table~\ref{tab:E1hb}
that our results are in reasonable agreement with experiment except for the case
$h_{b}(2P)\to \eta_{b}(2S)\gamma$ which is roughly a factor of $2$ lower. It is
worth to mention that the error is bigger in this case and the experimental
figure also seems to be higher than other theoretical
predictions~\cite{Godfrey:2002rp}.

There are four bottomonium states involved in the decay processes shown in
Table~\ref{tab:E1hb} which are still experimentally missing. These are the
$\eta_{b}(3S)$, $h_{b}(3P)$, $\eta_{b2}(1D)$ and $\eta_{b2}(2D)$. We have 
discussed already the $\eta_{b}(3S)$ state and we shall postpone for later 
on our discussion about the $1D$ and $2D$ states of the $\eta_{b2}$ meson. The 
$h_{b}(3P)$ meson has branching fractions of about $8\%$, $7\%$ and $13\%$ 
for its radiative decays into the $\eta_{b}(1S)$, $\eta_{b}(2S)$ and 
$\eta_{b}(3S)$ states, respectively. Since the radiative decay rates of the 
$\eta_{b}(3S)$ into the already observed $h_{b}(1P)$ and $h_{b}(2P)$ are, 
respectively, $0.0084$ and $2.60\,{\rm keV}$ (see Table~\ref{tab:E1etab}), the 
decay chain $h_{b}(3P)\to \eta_{b}(3S)\gamma\to h_{b}(2P)\gamma\gamma$ appears 
as the most suitable way for observing the $h_{b}(3P)$ and $\eta_{b}(3S)$ 
states. The branching fraction of the $h_{b}(3P)$ radiative decay into 
$\gamma\eta_{b2}(2D)$ is non negligible with a value of $5\%$ and can 
represent an opportunity to observe the $\eta_{b2}(2D)$ once the $h_{b}(3P)$ 
is established.

\begin{table}[!t]
\begin{center}
\scalebox{0.86}{\begin{tabular}{llrrc}
\hline
\hline
\tstrut
Initial state & Final state & $\Gamma_{\rm The.}$ & ${\cal B}_{\rm The.}$ 
& ${\cal B}_{\rm Exp.}$~\cite{PDG2014} \\
& & (keV) & $(\times10^{-2})$ & $(\times10^{-2})$ \\
\hline
\tstrut
$\chi_{b0}(1P)$ & $gg$                     & $2.00\,{\rm MeV}$ & $98.61$             & - \\
                & $\gamma\gamma$           & $0.12$            & $5.91\times10^{-3}$ & - \\
                & $\gamma\Upsilon(1S)$     & $28.07$           & $1.38$              & $1.76\pm0.30\pm0.18$ \\
                & total                    & $2.03\,{\rm MeV}$ & $100.00$            & - \\[2ex]
$\chi_{b1}(1P)$ & $q\bar{q}+g$             & $71.53$  & $66.73$  & - \\
                & $\gamma\Upsilon(1S)$     & $35.66$  & $33.27$  & $33.9\pm2.2$ \\
                & total                    & $107.19$ & $100.00$ & - \\[2ex]
$\chi_{b2}(1P)$ & $gg$                     & $83.69$             & $68.13$             & - \\
                & $\gamma\gamma$           & $3.08\times10^{-3}$ & $2.51\times10^{-3}$ & - \\
                & $\gamma\Upsilon(1S)$     & $39.15$             & $31.87$             & $19.1\pm1.2$ \\
                & $\gamma h_{b}(1P)$       & $8.88\times10^{-5}$ & $7.23\times10^{-5}$ & - \\
                & total                    & $122.84$            & $100.00$            & - \\[2ex]
$\chi_{b0}(2P)$ & $gg$                     & $2.37\,{\rm MeV}$   & $99.17$             & - \\
                & $\gamma\gamma$           & $0.14$              & $5.85\times10^{-3}$ & - \\
                & $\gamma\Upsilon(1S)$     & $5.44$              & $0.23$              & $0.9\pm0.6$ \\
                & $\gamma\Upsilon(2S)$     & $12.80$             & $0.54$              & $4.6\pm2.1$ \\
                & $\gamma\Upsilon(1D)$     & $0.74$              & $3.09\times10^{-2}$ & - \\
                & $\gamma h_{b}(1P)$       & $2.39\times10^{-3}$ & $9.99\times10^{-5}$ & - \\
                & $\pi\pi\chi_{b0}(1P)$    & $0.72$              & $3.01\times10^{-2}$ & - \\
                & $\pi\pi\chi_{b2}(1P)$    & $4.08\times10^{-5}$ & $1.71\times10^{-6}$ & - \\                
                & total                    & $2.39\,{\rm MeV}$   & $100.00$            & - \\[2ex]
$\chi_{b1}(2P)$ & $q\bar{q}+g$             & $106.14$            & $79.57$             & - \\
                & $\gamma\Upsilon(1S)$     & $9.13$              & $6.84$              & $9.2\pm0.8$ \\
                & $\gamma\Upsilon(2S)$     & $15.89$             & $11.91$             & $19.9\pm1.9$ \\
                & $\gamma\Upsilon(1D)$     & $0.41$              & $0.31$              & - \\
                & $\gamma\Upsilon_{2}(1D)$ & $1.26$              & $0.95$              & - \\
                & $\gamma h_{b}(1P)$       & $1.67\times10^{-4}$ & $1.25\times10^{-4}$ & - \\
                & $\pi\pi\chi_{b1}(1P)$    & $0.57$              & $0.43$              & $0.91\pm0.13$ \\
                & $\pi\pi\chi_{b2}(1P)$    & $1.94\times10^{-4}$ & $1.45\times10^{-4}$ & - \\
                & total                    & $133.40$            & $100.00$            & - \\[2ex]
$\chi_{b2}(2P)$ & $gg$                     & $104.26$            & $76.62$             & - \\
                & $\gamma\gamma$           & $3.84\times10^{-3}$ & $2.82\times10^{-3}$ & - \\
                & $\gamma\Upsilon(1S)$     & $11.38$             & $8.36$              & $7.0\pm0.7$ \\
                & $\gamma\Upsilon(2S)$     & $17.50$             & $12.86$             & $10.6\pm2.6$ \\
                & $\gamma\Upsilon(1D)$     & $2.09\times10^{-2}$ & $1.54\times10^{-2}$ & - \\
                & $\gamma\Upsilon_{2}(1D)$ & $0.35$              & $0.26$ & - \\
                & $\gamma\Upsilon_{3}(1D)$ & $2.06$              & $1.51$ & - \\
                & $\gamma h_{b}(1P)$       & $1.78\times10^{-3}$ & $1.31\times10^{-3}$ & - \\
                & $\gamma h_{b}(2P)$       & $2.86\times10^{-5}$ & $2.10\times10^{-5}$ & - \\
                & $\pi\pi\chi_{b0}(1P)$    & $8.49\times10^{-6}$ & $6.24\times10^{-6}$ & - \\
                & $\pi\pi\chi_{b1}(1P)$    & $6.06\times10^{-4}$ & $4.45\times10^{-4}$ & - \\
                & $\pi\pi\chi_{b2}(1P)$    & $0.49$              & $0.36$ & $0.51\pm0.09$ \\                
                & total                    & $136.07$            & $100.00$ & - \\
\hline
\hline
\end{tabular}}
\caption{\label{tab:E1chibJ_v1} Decay widths and branching fractions of annihilation 
rates, radiative decays and hadronic transitions for the $\chi_{bJ}$ states.
The experimental data are from Ref.~\cite{PDG2014}.}
\end{center}
\end{table}

\begin{table}[!t]
\begin{center}
\begin{tabular}{llrr}
\hline
\hline
\tstrut
Initial state & Final state & $\Gamma_{\rm The.}$ & ${\cal B}_{\rm The.}$ \\
& & (keV) & $(\times10^{-2})$ \\
\hline
\tstrut
$\chi_{b0}(3P)$ & $gg$                     & $2.46\,{\rm MeV}$   & $99.26$             \\
                & $\gamma\gamma$           & $0.15$              & $6.06\times10^{-3}$ \\
                & $\gamma\Upsilon(1S)$     & $1.99$              & $8.04\times10^{-2}$ \\
                & $\gamma\Upsilon(2S)$     & $2.99$              & $0.12$              \\
                & $\gamma\Upsilon(3S)$     & $8.50$              & $0.34$              \\
                & $\gamma\Upsilon(1D)$     & $3.59\times10^{-2}$ & $1.45\times10^{-3}$ \\
                & $\gamma\Upsilon(2D)$     & $3.50$              & $0.14$              \\
                & $\pi\pi\chi_{b0}(1P)$    & $1.16$              & $4.69\times10^{-2}$ \\
                & $\pi\pi\chi_{b2}(1P)$    & $4.28\times10^{-3}$ & $1.73\times10^{-4}$ \\
                & total                    & $2.47\,{\rm MeV}$   & $100.00$            \\[2ex]
$\chi_{b1}(3P)$ & $q\bar{q}+g$             & $124.53$            & $83.59$             \\
                & $\gamma\Upsilon(1S)$     & $4.17$              & $2.80$              \\
                & $\gamma\Upsilon(2S)$     & $4.58$              & $3.07$              \\
                & $\gamma\Upsilon(3S)$     & $9.62$              & $6.46$              \\
                & $\gamma\Upsilon(1D)$     & $4.80\times10^{-2}$ & $3.22\times10^{-2}$ \\
                & $\gamma\Upsilon(2D)$     & $1.26$              & $0.85$              \\
                & $\gamma\Upsilon_{2}(1D)$ & $0.11$              & $7.38\times10^{-2}$ \\
                & $\gamma\Upsilon_{2}(2D)$ & $3.34$              & $2.24$              \\
                & $\pi\pi\chi_{b1}(1P)$    & $1.32$              & $0.89$              \\
                & $\pi\pi\chi_{b2}(1P)$    & $4.55\times10^{-3}$ & $3.05\times10^{-3}$ \\
                & total                    & $148.98$            & $100.00$            \\[2ex]
$\chi_{b2}(3P)$ & $gg$                     & $111.45$            & $79.56$             \\
                & $\gamma\gamma$           & $4.10\times10^{-3}$ & $2.93\times10^{-3}$ \\
                & $\gamma\Upsilon(1S)$     & $5.65$              & $4.03$              \\
                & $\gamma\Upsilon(2S)$     & $5.62$              & $4.01$              \\
                & $\gamma\Upsilon(3S)$     & $10.38$             & $7.41$              \\
                & $\gamma\Upsilon(1D)$     & $3.38\times10^{-3}$ & $2.41\times10^{-3}$ \\
                & $\gamma\Upsilon(2D)$     & $0.18$              & $0.13$              \\
                & $\gamma\Upsilon_{2}(1D)$ & $4.41\times10^{-2}$ & $3.15\times10^{-2}$ \\
                & $\gamma\Upsilon_{2}(2D)$ & $0.79$              & $0.56$              \\
                & $\gamma\Upsilon_{3}(1D)$ & $0.21$              & $0.15$              \\
                & $\gamma\Upsilon_{3}(2D)$ & $4.16$              & $2.97$              \\
                & $\pi\pi\chi_{b0}(1P)$    & $1.88\times10^{-3}$ & $1.34\times10^{-3}$ \\
                & $\pi\pi\chi_{b1}(1P)$    & $4.37\times10^{-3}$ & $3.12\times10^{-3}$ \\
                & $\pi\pi\chi_{b2}(1P)$    & $1.59$              & $1.14$              \\
                & total                    & $140.09$            & $100.00$            \\
\hline
\hline
\end{tabular}
\caption {\label{tab:E1chibJ_v2} (Continuation) Decay widths and branching 
fractions of annihilation rates, radiative decays and hadronic transitions 
for the $\chi_{bJ}$ states. There is no experimental data available.}
\end{center}
\end{table}

Tables~\ref{tab:E1chibJ_v1} and~\ref{tab:E1chibJ_v2} show E1 and M1 radiative 
decays of the $\chi_{bJ}(nP)$ states with $J=0,\,1,\,2$ and $n=1,\,2,\,3$. 
These Tables also show some hadronic transitions (the spin-nonflip $\pi\pi$ 
transitions) and the annihilation rates into gluons. As one can see, the 
available experimental data is scarce and mostly related with the E1 radiative 
decays associated with the decay channels through which the $\chi_{bJ}(1P)$ and 
$\chi_{bJ}(2P)$ were discovered in the early eighties. Our theoretical results 
are in reasonable agreement with the experimental figures except for the case 
of the $\chi_{b0}(2P)$, in which our predictions of the branching fractions are 
much smaller than the experimental data. There could be two reasons for this. 
The first one could be related with an overestimation of the decay rate for the 
annihilation into gluons of the $\chi_{b0}(2P)$ state. However, This should 
have been reflected also in the $\chi_{b0}(1P)$ state and even more strongly 
because the smaller number of open decay channels. This is not the case as 
reflected in Table~\ref{tab:E1chibJ_v1}. Therefore, if we are overestimating 
the $\chi_{b0}(2P)$ annihilation into gluons, this cannot be in a dramatic way. 
The second possibility could be an error on the experimental measurements which 
have, up to now, uncertainties in the order of the fifty percent.

One can inferred from the Tables~\ref{tab:E1chibJ_v1} and~\ref{tab:E1chibJ_v2}
that the $\chi_{b1}(1P,2P,3P)$ and $\chi_{b2}(1P,2P,3P)$ mesons have total decay
widths of around $100-150\,{\rm keV}$ whereas the $\chi_{b0}(1P,2P,3P)$ states 
have total decay widths of about $2-2.5\,{\rm MeV}$. The contribution of the 
decay rate into gluons is $99\%$ in the case of the $\chi_{b0}$ states 
compared with the $70-80\%$ for the $\chi_{b1}$ and $\chi_{b2}$ mesons.

Special attention deserves Table~\ref{tab:E1chibJ_v2} in which we have 
collected the decay properties of the spin-triplet $3P$-wave states. We hope 
that the theoretical data shown in Table~\ref{tab:E1chibJ_v2} help 
experimentalists in carrying out an intensive study of them. Some of their 
radiative decays are dominant, with rates in the order of few keV. Therefore, 
these transitions still seem to be the best way for disentangling the fine mass 
splittings. The spin-nonflip $\pi\pi$ transitions 
$\chi_{b0}(3P)\to\chi_{b0}(1P)$, $\chi_{b1}(3P)\to\chi_{b1}(1P)$ and 
$\chi_{b2}(3P)\to\chi_{b2}(1P)$ have decay rates lower but in the same order of 
magnitude, around $1-2\,{\rm keV}$, and thus can be also used for studying the 
$3P$ spin-triplet states.

We finish this Section calling the attention of the reader to the fact that our 
prediction of the spin-nonflip $\pi\pi$ transitions between $\chi_{bJ}$ states 
follow a particular path: the decay rates between states with $J_{i}=J_{f}$ are 
orders of magnitude higher than those with $J_{i}\neq J_{f}$. This is so 
because the first type of transitions goes through the term with the coefficient 
$C_{1}$ of Eq.~(\ref{HofE1E1}) whereas the second type of transitions involve 
only the term of Eq.~(\ref{HofE1E1}) with constant $C_{2}$ that is much 
smaller. The constant $C_{1}$ is fitted through the 
$\Upsilon(2S)\to\pi\pi\Upsilon(1S)$ decay whereas the coefficient $C_{2}$ is 
fitted through the transition $\Upsilon_{2}(1D)\to\pi\pi\Upsilon(1S)$. It is 
remarkable that, as seen in Table~\ref{tab:E1chibJ_v1}, we obtain good 
agreement in those cases in which experimental data are available.

%%%%%%%%%%%%%%%%%%%%%%%%%%%%%%%%%%%%%%%%%%%%%%%%%%%%%%%%%%%%%%%%%%%%%%%%%%%%%%%%

\subsection{The $S$-wave $\Upsilon$ levels}

The success of QCD-inspired potentials is due largely to the fruitful
description and prediction of the properties of the the $S$-wave $\psi$ and
$\Upsilon$ states. One can see in Table~\ref{tab:predmassesbb} that the masses
of $\Upsilon(1S)$, $\Upsilon(2S)$ and $\Upsilon(3S)$ located experimentally at
$9.46$, $10.02$ and $10.35\,{\rm GeV}$ are reasonably well reproduce in our
quark model: $9.50$, $10.02$ and $10.35\,{\rm GeV}$, respectively.

We show in Table~\ref{tab:E1Supsilon} the decay widths and branching fractions 
of annihilation rates, radiative decays and hadronic transitions for the 
$\Upsilon(1S)$, $\Upsilon(2S)$ and $\Upsilon(3S)$ states. In order to avoid 
unnecessary uncertainties in the calculated branching fractions, we use in these 
cases the total decay widths available in PDG~\cite{PDG2014}.

\begin{table}[!t]
\begin{center}
\scalebox{0.80}{\begin{tabular}{llrrc}
\hline
\hline
\tstrut
Initial state & Final state & $\Gamma_{\rm The.}$ & ${\cal B}_{\rm The.}$ &
${\cal B}_{\rm Exp.}$ \\
& & (keV) & $(\times10^{-2})$ & $(\times10^{-2})$ \\
\hline
\tstrut
$\Upsilon(1S)$ & $e^{+}e^{-}$         & $0.71$              & $1.31$              & $2.38\pm0.11$ \\
               & $3g$                 & $41.63$             & $77.06$             & $81.7\pm0.7$ \\
               & $\gamma gg$          & $0.79$              & $1.46$              & $2.2\pm0.6$ \\
               & $3\gamma$            & $3.44\times10^{-6}$ & $6.37\times10^{-6}$ & - \\ 
               & $\gamma\eta_{b}(1S)$ & $9.34\times10^{-3}$ & $1.73\times10^{-2}$ & - \\[2ex]
$\Upsilon(2S)$ & $e^{+}e^{-}$          & $0.37$              & $1.16$              & $1.91\pm0.16$ \\
               & $3g$                  & $24.25$             & $75.83$             & $58.8\pm1.2$ \\
               & $\gamma gg$           & $0.46$              & $1.44$              & $8.8\pm1.1$ \\
               & $3\gamma$             & $2.00\times10^{-6}$ & $6.25\times10^{-6}$ & - \\ 
               & $\gamma\chi_{b0}(1P)$ & $1.09$              & $3.41$              & $3.8\pm0.4$ \\
               & $\gamma\chi_{b1}(1P)$ & $1.84$              & $5.75$              & $6.9\pm0.4$ \\
               & $\gamma\chi_{b2}(1P)$ & $2.08$              & $6.50$              & $7.15\pm0.35$ \\
               & $\gamma\eta_{b}(1S)$  & $5.65\times10^{-2}$ & $0.18$              & $0.11\pm0.04^{+0.07}_{-0.05}$~\cite{Lees:2011mx} \\
               & $\gamma\eta_{b}(2S)$  & $5.80\times10^{-4}$ & $1.81\times10^{-3}$ & - \\
               & $\pi\pi\Upsilon(1S)$  & $8.57$              & $26.80$             & $26.45\pm0.48$ \\[2ex]
$\Upsilon(3S)$ & $e^{+}e^{-}$          & $0.27$              & $1.33$              & $2.18\pm0.20$ \\
               & $3g$                  & $18.76$             & $92.32$             & $35.7\pm2.6$ \\
               & $\gamma gg$           & $0.36$              & $1.77$              & $0.97\pm0.18$ \\
               & $3\gamma$             & $1.55\times10^{-6}$ & $7.63\times10^{-6}$ & - \\ 
               & $\gamma\chi_{b0}(1P)$ & $0.15$              & $0.74$              & $0.27\pm0.04$ \\
               & $\gamma\chi_{b1}(1P)$ & $0.16$              & $0.79$              & $0.09\pm0.05$ \\
               & $\gamma\chi_{b2}(1P)$ & $8.27\times10^{-2}$ & $0.41$              & $0.99\pm0.13$ \\
               & $\gamma\chi_{b0}(2P)$ & $1.21$              & $5.96$              & $5.9\pm0.6$ \\
               & $\gamma\chi_{b1}(2P)$ & $2.13$              & $10.48$             & $12.6\pm1.2$ \\
               & $\gamma\chi_{b2}(2P)$ & $2.56$              & $12.60$             & $13.1\pm1.6$ \\
               & $\gamma\eta_{b}(1S)$  & $5.70\times10^{-2}$ & $0.28$              & $0.058\pm0.016^{+0.014}_{-0.016}$~\cite{Lees:2011mx} \\
               & $\gamma\eta_{b}(2S)$  & $1.10\times10^{-2}$ & $5.41\times10^{-2}$ & $<0.062$ \\
               & $\gamma\eta_{b}(3S)$  & $6.58\times10^{-4}$ & $3.24\times10^{-3}$ & - \\
               & $\pi\pi\Upsilon(1S)$  & $1.77$              & $8.71$              & $ 6.57\pm0.15$ \\
               & $\pi\pi\Upsilon(2S)$  & $0.42$              & $2.07$              & $4.67\pm0.23$ \\
\hline
\hline
\end{tabular}}
\caption{\label{tab:E1Supsilon} Decay widths and branching 
fractions of annihilation rates, radiative decays and hadronic transitions 
for the $\Upsilon(1S)$, $\Upsilon(2S)$ and $\Upsilon(3S)$ states. The 
experimental data are from Refs.~\cite{PDG2014,Lees:2011mx}. In this case, we 
have calculated the branching fractions using the experimental total decay widths of PDG2014.}
\end{center}
\end{table}

The annihilation rates of the $\Upsilon(1S)$ state are slightly lower than the 
experimental data but in reasonable agreement. Unfortunately, there is no 
experimental data associated with the only radiative decay of the 
$\Upsilon(1S)$ state. The $\Upsilon(1S)\to \gamma\eta_{b}(1S)$ is an M1 
transition and thus it is suppressed with respect the E1 decays and much more 
difficult to measure. This can be observed in the cases of the $\Upsilon(2S)$ 
and $\Upsilon(3S)$ where transitions into $\eta_{b}$ states are orders of 
magnitude smaller than the transitions into $\chi_{bJ}$ states. Using 
pNRQCD~\cite{Brambilla:2005zw}, the authors of Ref.~\cite{Pineda:2013lta} have 
recently computed the decay rate of the $\Upsilon(1S)\to \gamma\eta_{b}(1S)$ 
transition reporting a value of $(0.01518\pm0.00051)\,{\rm keV}$, which is 
higher than our quark model prediction of $\sim0.0093\,{\rm keV}$. This decay 
rate is extremely sensitive to the masses of the $\Upsilon(1S)$ and 
$\eta_{b}(1S)$ mesons. If we compute the decay width using the masses predicted 
by pNRQCD, our result is $0.014\,{\rm keV}$ and agrees with the pNRQCD value.

As one can see in Table~\ref{tab:E1Supsilon}, all branching fractions of the 
$\Upsilon(2S)$ state which are related with annihilation, E1 and M1 radiative 
decays and even the spin-nonflip $\pi\pi$ transitions are in reasonable 
agreement with the experimental data. In Ref.~\cite{Lees:2011mx}, Lees {\it et
al.} have performed a study of radiative transitions between bottomonium states 
with a huge amount of events recorded by the BaBar detector at the PEP-II 
B-factory at SLAC. Among their measurements, a value of the $\Upsilon(2S)\to 
\gamma\eta_{b}(1S)$ decay rate is reported and agrees, within errors, with our 
theoretical figure (see Table~\ref{tab:E1Supsilon}). However, it is important 
to emphasize that this decay rate is far from being to be well established. On 
the experimental side, the PDG of 2014 collects the branching fraction 
$(3.9\pm1.5)\times10^{-4}$ from an early measurement of the BaBar 
Collaboration~\cite{Aubert:2009as} despite of having the measurement of $2011$ 
reported in Ref.~\cite{Lees:2011mx}. On the theoretical side, there are values 
in lattice NRQCD~\cite{Hughes:2015dba}, $(5.4\pm1.8)\times10^{-4}$, in 
continuum pNRQCD~\cite{Pineda:2013lta}, $(1.88\pm8.34)\times10^{-4}$, and within 
quark models ranging from $0.05\times10^{-4}$ to $18\times10^{-4}$.

The nice agreement with experimental data seems to change for the 
$\Upsilon(3S)$ state. While we reproduce the branching fractions for the 
radiative decays of the $\Upsilon(3S)$ into $\chi_{bJ}(2P)$ states and also for 
the $2$-pion decays into $\Upsilon(1S)$ and $\Upsilon(2S)$ states, we are 
only able to give the correct order of magnitude for the E1 radiative decays of 
the $\Upsilon(3S)$ into $\chi_{bJ}(1P)$ states and the annihilation rate into 
gluons seems to be predicted much larger than the experimental figure.

\begin{table}[!t]
\begin{center}
\scalebox{0.90}{\begin{tabular}{cccrrrrr}
\hline
\hline
\multicolumn{3}{c}{Decay chain} & ${\cal B}_{1}$ & ${\cal B}_{2}$ & ${\cal
B}_{3}$ & ${\cal B}_{\rm The.}$ & ${\cal B}_{\rm Exp.}$~\cite{Lees:2014qea} \\
& & & $(\%)$ & $(\%)$ & $(\%)$ & $(10^{-4})$ & $(10^{-4})$ \\
\hline
\tstrut
$2^{3}S_1$ & $\to 1^{3}P_{0}$ & $\to 1^{3}S_{1}$ & $3.41$ & $1.38$  & $2.48$ & $0.12$ & $0.29^{+0.17+0.01}_{-0.14-0.08}$ \\
$2^{3}S_1$ & $\to 1^{3}P_{1}$ & $\to 1^{3}S_{1}$ & $5.75$ & $33.27$ & $2.48$ & $4.74$ & $6.86^{+0.47+0.44}_{-0.45-0.35}$ \\
$2^{3}S_1$ & $\to 1^{3}P_{2}$ & $\to 1^{3}S_{1}$ & $6.50$ & $31.87$ & $2.48$ & $5.14$ & $3.63^{+0.36+0.18}_{-0.34-0.19}$ \\[2ex]
$3^{3}S_1$ & $\to 2^{3}P_{0}$ & $\to 2^{3}S_{1}$ & $5.96$  & $0.54$  & $1.93$ & $0.062$ & $0.66^{+0.49+0.20}_{-0.40-0.03}$ \\
$3^{3}S_1$ & $\to 2^{3}P_{1}$ & $\to 2^{3}S_{1}$ & $10.48$ & $11.91$ & $1.93$ & $2.41$  & $4.95^{+0.75+1.01}_{-0.70-0.24}$ \\
$3^{3}S_1$ & $\to 2^{3}P_{2}$ & $\to 2^{3}S_{1}$ & $12.60$ & $12.86$ & $1.93$ & $3.13$  & $3.22^{+0.58+0.16}_{-0.53-0.71}$ \\[2ex]
$3^{3}S_1$ & $\to 2^{3}P_{0}$ & $\to 1^{3}S_{1}$ & $5.96$  & $0.23$ & $2.48$ & $0.034$ & $0.17^{+0.15+0.01}_{-0.14-0.12}$ \\
$3^{3}S_1$ & $\to 2^{3}P_{1}$ & $\to 1^{3}S_{1}$ & $10.48$ & $6.84$ & $2.48$ & $1.78$  & $3.52^{+0.28+0.17}_{-0.27-0.18}$ \\
$3^{3}S_1$ & $\to 2^{3}P_{2}$ & $\to 1^{3}S_{1}$ & $12.60$ & $8.36$ & $2.48$ & $2.61$  & $1.95^{+0.22+0.10}_{-0.21-0.16}$ \\[2ex]
$3^{3}S_1$ & $\to 1^{3}P_{0}$ & $\to 1^{3}S_{1}$ & $0.74$ & $1.38$  & $2.48$ & $0.025$ & - \\
$3^{3}S_1$ & $\to 1^{3}P_{1}$ & $\to 1^{3}S_{1}$ & $0.79$ & $33.27$ & $2.48$ & $0.65$  & $1.16^{+0.78+0.14}_{-0.67-0.16}$ \\
$3^{3}S_1$ & $\to 1^{3}P_{2}$ & $\to 1^{3}S_{1}$ & $0.41$ & $31.87$ & $2.48$ & $0.32$  & $4.68^{+0.99}_{-0.92}\pm0.37$ \\
\hline
\hline
\end{tabular}}
\caption{\label{tab:chainSupsilon} Radiative decay chains of the
$\Upsilon(2S)$ and $\Upsilon(3S)$ states involving the $\chi_{bJ}(1P,2P)$
mesons. The branching fractions are ${\cal B}_{1}={\cal B}(n^{3}S_{1}\to
m^{3}P_{J}+\gamma)$, ${\cal B}_{2}={\cal B}(m^{3}P_{J}\to n'^{3}S_{1}+\gamma)$,
and ${\cal B}_{3}={\cal B}(n'^{3}S_{1}\to \mu^{+}\mu^{-})$. For the theoretical
calculation, we take the branching fraction ${\cal B}_{3}$ from PDG2014. The
experimental data is taken from Ref.~\cite{Lees:2014qea}.}
\end{center}
\end{table}

\begin{table}[!t]
\begin{center}
\begin{tabular}{ccccc}
\hline
\hline
\tstrut
$\Upsilon(nS)$ & Theory & BaBar~\cite{Aubert:2008ab} & Belle~\cite{Chen:2008xia}
& PDG2014~\cite{PDG2014} \\
\hline
\tstrut
$5S$ & $10818$ & $10876\pm2$ & $10879\pm3$ & $10876\pm11$ \\
$6S$ & $10995$ & $10996\pm2$ & -           & $11019\pm8$ \\
\hline
\hline
\end{tabular}
\caption{\label{tab:newpar56S} New masses, in MeV, reported by the BaBar and
Belle Collaborations for the $\Upsilon(5S)$ and $\Upsilon(6S)$ states. We
compare with our theoretical results and the current PDG2014 values.}
\end{center}
\end{table}

\begin{table}[!t]
\begin{center}
\scalebox{0.835}{\begin{tabular}{llrrc}
\hline
\hline
\tstrut
Initial state & Final state & $\Gamma_{\rm The.}$ & ${\cal B}_{\rm The.}$ &
${\cal B}_{\rm Exp.}$ \\
& & (keV) & $(\times10^{-2})$ & $(\times10^{-2})$ \\
\hline
\tstrut
$\Upsilon(4S)$ & $e^{+}e^{-}$          & $0.21$              & $1.02\times10^{-3}$ & $(1.57\pm0.08)\times10^{-3}$ \\
               & $3g$                  & $15.58$             & $7.60\times10^{-2}$ & - \\
               & $\gamma gg$           & $0.30$              & $1.46\times10^{-3}$ & - \\
               & $3\gamma$             & $1.29\times10^{-6}$ & $6.29\times10^{-9}$ & - \\
               & $\gamma\chi_{b0}(1P)$ & $5.88\times10^{-2}$ & $2.87\times10^{-4}$ & - \\
               & $\gamma\chi_{b1}(1P)$ & $4.74\times10^{-2}$ & $2.31\times10^{-4}$ & - \\
               & $\gamma\chi_{b2}(1P)$ & $1.20\times10^{-2}$ & $5.85\times10^{-5}$ & - \\               
               & $\gamma\chi_{b0}(2P)$ & $0.17$              & $8.29\times10^{-4}$ & - \\
               & $\gamma\chi_{b1}(2P)$ & $0.18$              & $8.78\times10^{-4}$ & - \\
               & $\gamma\chi_{b2}(2P)$ & $0.11$              & $5.37\times10^{-4}$ & - \\               
               & $\gamma\chi_{b0}(3P)$ & $0.61$              & $2.98\times10^{-3}$ & - \\
               & $\gamma\chi_{b1}(3P)$ & $1.17$              & $5.71\times10^{-3}$ & - \\
               & $\gamma\chi_{b2}(3P)$ & $1.45$              & $7.07\times10^{-3}$ & - \\               
               & $\gamma\eta_{b}(1S)$  & $4.98\times10^{-2}$ & $2.43\times10^{-4}$ & - \\
               & $\gamma\eta_{b}(2S)$  & $1.24\times10^{-2}$ & $6.05\times10^{-5}$ & - \\               
               & $\gamma\eta_{b}(3S)$  & $3.88\times10^{-3}$ & $1.89\times10^{-5}$ & - \\
               & $\pi\pi\Upsilon(1S)$  & $6.02$              & $2.94\times10^{-2}$ & $(1.22\pm0.09)\times10^{-2}$ \\
               & $\pi\pi\Upsilon(2S)$  & $0.24$              & $1.17\times10^{-3}$ & $(1.29\pm0.20)\times10^{-2}$ \\[2ex]
$\Upsilon(10860)$ & $e^{+}e^{-}$          & $0.18$              & $3.27\times10^{-4}$ & $(5.6\pm3.1)\times10^{-4}$ \\
                  & $3g$                  & $13.33$             & $2.42\times10^{-2}$ & - \\
                  & $\gamma gg$           & $0.25$              & $4.55\times10^{-4}$ & - \\
                  & $3\gamma$             & $1.10\times10^{-6}$ & $2.00\times10^{-9}$ & - \\
                  & $\gamma\chi_{b0}(1P)$ & $6.85\times10^{-2}$ & $1.25\times10^{-4}$ & - \\
                  & $\gamma\chi_{b1}(1P)$ & $6.29\times10^{-2}$ & $1.14\times10^{-4}$ & - \\
                  & $\gamma\chi_{b2}(1P)$ & $2.26\times10^{-2}$ & $4.11\times10^{-5}$ & - \\               
                  & $\gamma\chi_{b0}(2P)$ & $0.22$              & $4.00\times10^{-4}$ & - \\
                  & $\gamma\chi_{b1}(2P)$ & $0.26$              & $4.73\times10^{-4}$ & - \\
                  & $\gamma\chi_{b2}(2P)$ & $0.18$              & $3.27\times10^{-4}$ & - \\               
                  & $\gamma\chi_{b0}(3P)$ & $0.80$              & $1.45\times10^{-3}$ & - \\
                  & $\gamma\chi_{b1}(3P)$ & $1.35$              & $2.45\times10^{-3}$ & - \\
                  & $\gamma\chi_{b2}(3P)$ & $1.42$              & $2.58\times10^{-3}$ & - \\               
                  & $\gamma\eta_{b}(1S)$  & $5.97\times10^{-2}$ & $1.09\times10^{-4}$ & - \\
                  & $\gamma\eta_{b}(2S)$  & $2.17\times10^{-2}$ & $3.95\times10^{-5}$ & - \\               
                  & $\gamma\eta_{b}(3S)$  & $1.37\times10^{-2}$ & $2.49\times10^{-5}$ & - \\
                  & $\pi\pi\Upsilon(1S)$  & $12.82$             & $2.33\times10^{-2}$ & $0.80\pm0.09$ \\
                  & $\pi\pi\Upsilon(2S)$  & $17.89$             & $3.25\times10^{-2}$ & $1.17\pm0.20$ \\
                  & $\pi\pi\Upsilon(3S)$  & $5.56$              & $1.01\times10^{-2}$ & $0.72^{+0.29}_{-0.26}$ \\
                  & $\pi\pi\Upsilon(4S)$  & $8.54\times10^{-2}$ & $1.55\times10^{-4}$ & - \\[2ex]
$\Upsilon(11020)$ & $e^{+}e^{-}$            & $0.15$              & $1.90\times10^{-4}$ & $(1.6\pm0.5)\times10^{-4}$ \\
                  & $3g$                    & $11.57$             & $1.46\times10^{-2}$ & - \\
                  & $\gamma gg$             & $0.22$              & $2.78\times10^{-4}$ & - \\
                  & $3\gamma$               & $9.56\times10^{-7}$ & $1.21\times10^{-9}$ & - \\
                  & $\gamma\chi_{b0}(1P)$   & $4.08\times10^{-2}$ & $5.16\times10^{-5}$ & - \\
                  & $\gamma\chi_{b1}(1P)$   & $3.32\times10^{-2}$ & $4.20\times10^{-5}$ & - \\
                  & $\gamma\chi_{b2}(1P)$   & $8.12\times10^{-3}$ & $1.03\times10^{-5}$ & - \\               
                  & $\gamma\chi_{b0}(2P)$   & $0.11$              & $1.39\times10^{-4}$ & - \\
                  & $\gamma\chi_{b1}(2P)$   & $0.11$              & $1.39\times10^{-4}$ & - \\
                  & $\gamma\chi_{b2}(2P)$   & $5.12\times10^{-2}$ & $6.48\times10^{-5}$ & - \\               
                  & $\gamma\chi_{b0}(3P)$   & $0.26$              & $3.29\times10^{-4}$ & - \\
                  & $\gamma\chi_{b1}(3P)$   & $0.35$              & $4.43\times10^{-4}$ & - \\
                  & $\gamma\chi_{b2}(3P)$   & $0.27$              & $3.42\times10^{-4}$ & - \\               
                  & $\gamma\eta_{b}(1S)$    & $5.08\times10^{-2}$ & $6.43\times10^{-5}$ & - \\
                  & $\gamma\eta_{b}(2S)$    & $1.87\times10^{-2}$ & $2.37\times10^{-5}$ & - \\               
                  & $\gamma\eta_{b}(3S)$    & $1.15\times10^{-2}$ & $1.46\times10^{-5}$ & - \\
                  & $\pi\pi\Upsilon(1S)$    & $276.20$            & $0.35$              & - \\
                  & $\pi\pi\Upsilon(2S)$    & $6.32$              & $8.00\times10^{-3}$ & - \\
                  & $\pi\pi\Upsilon(3S)$    & $38.81$             & $4.91\times10^{-2}$ & - \\
                  & $\pi\pi\Upsilon(4S)$    & $1.29$              & $1.63\times10^{-3}$ & - \\
\hline
\hline
\end{tabular}}
\caption{\label{tab:E1Supsilon_v2} Decay widths and branching 
fractions of annihilation rates, radiative decays and hadronic transitions 
for the $\Upsilon(4S)$, $\Upsilon(10860)$ and $\Upsilon(11020)$ states. The 
experimental data are from Refs.~\cite{PDG2014,Lees:2011mx}. In this case, we 
have calculated the branching fractions using the experimental total decay 
widths of PDG2014.}
\end{center}
\end{table}

In December of $2014$ the BaBar Collaboration published an experimental
work~\cite{Lees:2014qea} in which $121$ million of $\Upsilon(3S)$ and $98$
million of $\Upsilon(2S)$ mesons were used to perform a study of radiative
transitions involving the $\chi_{bJ}(1P,2P)$ states. This work includes the best
observational significance of some transitions and provide the most up-to-date
derived branching fractions in the bottomonium system. 
Table~\ref{tab:chainSupsilon} summarizes their primary results and compares with
our theoretical values. Our results are compatible within experimental errors in 
most of the cases but some discrepancies are also found.

Above the $B\bar{B}$ threshold, there are three more states well established in 
the PDG with quantum number $J^{PC}=1^{--}$. They are the so-called 
$\Upsilon(4S)$, $\Upsilon(10860)$ and $\Upsilon(11020)$, being the last two 
natural candidates for the $\Upsilon(5S)$ and $\Upsilon(6S)$, respectively. The 
$B$-factories scanned again the energy range above open-bottom threshold. The 
BaBar Collaboration~\cite{Aubert:2008ab} performed a comprehensive scan between 
$10.54$ and $11.2\,{\rm GeV}$, followed by an eight-point scan in the proximity 
of the $\Upsilon(6S)$ peak. The Belle Collaboration~\cite{Chen:2008xia} 
acquired nine points over $10.80-11.02\,{\rm GeV}$, as well as spread over 
seven additional points more focused on the $\Upsilon(5S)$ peak. Both scans 
suggest that the simple Breit-Wigner parametrization, previously used to model 
the peaks observed in the CLEO~\cite{Besson:1984bd} and 
CUSB~\cite{Lovelock:1985nb} scans, is not good enough for the description of 
the complex dynamics in the proximity of the $B^{(\ast)}\bar{B}^{(\ast)}$ and 
$B_s^{(\ast)}\bar{B_s}^{(\ast)}$ thresholds. The new data points on 
$R_{b}=\sigma(b\bar{b})/\sigma(\mu\mu)$ are better modeled assuming a flat 
$b\bar{b}$ continuum contribution which interferes constructively with the 
$\Upsilon(5S)$ and $\Upsilon(6S)$ Breit-Wigner resonances, and a second flat 
contribution which adds incoherently. Such fits alter the PDG results on the 
$\Upsilon(5S)$ and $\Upsilon(6S)$ peaks. Table~\ref{tab:newpar56S} compares the 
theoretical prediction with the new parameters reported by BaBar and Belle, but 
also with the PDG's values.

Table~\ref{tab:E1Supsilon_v2} shows the decay widths and branching fractions of 
annihilation rates, radiative decays and hadronic transitions for the 
$\Upsilon(4S)$, $\Upsilon(10860)$ and $\Upsilon(11020)$ states. The 
experimental data are from Refs.~\cite{PDG2014,Lees:2011mx}. We have 
again used the experimental total decay widths reported by PDG~\cite{PDG2014} 
in order to calculate the theoretical branching fractions. The theoretical 
total decay widths can be found on Table~\ref{tab:strongUps} which agrees well 
with the experimental values for the $\Upsilon(4S)$ and $\Upsilon(11020)$ 
states. Since the experimental data is scarce, we can only comment interesting 
features of our theoretical results. The $\Upsilon(nS)$ with $n=4,\,5,\,6$ have 
radiative decays into the $\chi_{bJ}(1P,2P,3P)$ states whose widths go from 
$0.01$ to $1.5\,{\rm keV}$. Their M1 radiative decays into the $\eta_{b}$ 
states are $2-3$ orders of magnitude smaller.

We compare in Table~\ref{tab:E1Supsilon_v2} the spin-nonflip $\pi\pi$ hadronic 
transitions of the $\Upsilon(4S)$, $\Upsilon(10860)$ and $\Upsilon(11020)$ 
states for which experimental data are available. One can see that our values 
agree reasonably well with the experimental ones except in the case of the 
$\Upsilon(10860)$. We have seen in Refs.~\cite{Segovia:2014mca,Segovia:2015raa} 
that the presence of hybrid mesons close in mass to conventional quarkonium 
states leads to large enhancements in some hadronic transition decay rates. We 
do not find any hybrid state around the $\Upsilon(10860)$ resonance. Therefore, 
other mechanism is needed in order to explain the large $\pi\pi$ decay rates 
observed experimentally. The authors of Ref.~\cite{Ali:2009es} have recently 
analyzed the Belle data on the cross section of the process $e^{+}e^{-} \to 
\Upsilon(nS) \pi\pi$ with $n=1,\,2$ and around the $\Upsilon(10860)$ energy 
region. They found that the experimental data is compatible with a tetraquark 
interpretation for the $\Upsilon(10860)$. Further studies are needed to 
understand the anomalous $\pi\pi$ decay widths of the $\Upsilon(10860)$.

\begin{table}[!t]
\begin{center}
\begin{tabular}{lllrrc}
\hline
\hline
\tstrut
Meson & State & Channel & $\Gamma_{^{3}P_{0}}$ & ${\mathcal B}_{^3P_0}$ &
${\cal B}_{\rm Exp.}$~\cite{PDG2014} \\
\hline
\tstrut
$\Upsilon(4S)$ & $4^{3}S_{1}$ & $B^{+}B^{-}$       & $10.41$ & $50.54$  & $51.3\pm0.6$ \\
               &              & $B^{0}\bar{B}^{0}$ & $10.18$ & $49.46$  & $48.7\pm0.6$ \\
               &              & $BB$               & $20.59$ & $100.00$ & $>96$ \\
$20.5\pm2.5$   &              & total              & $20.59$ & $100.00$ & - \\[2ex]
$\Upsilon(10860)$ & $5^{3}S_{1}$ & $BB$                           & $6.22$  & $22.29$  & $5.5\pm1.0$ \\
                  &              & $BB^{\ast}$                    & $11.83$ & $42.41$  & $13.7\pm1.6$ \\
                  &              & $B^{\ast}B^{\ast}$             & $0.09$  & $0.32$   & $38.1\pm3.4$ \\
                  &              & $B_{s}B_{s}$                   & $0.96$  & $3.45$   & $0.5\pm0.5$ \\
                  &              & $B_{s}B_{s}^{\ast}$            & $1.15$  & $4.11$   & $1.5\pm0.7$ \\
                  &              & $B_{s}^{\ast}B_{s}^{\ast}$     & $7.65$  & $27.42$  & $17.9\pm2.8$ \\
                  &              & $B_{s}^{(\ast)}B_{s}^{(\ast)}$ & $9.76$  & $34.98$  & $19.9\pm3.0$ \\
$55\pm28$         &              & total                          & $27.89$ & $100.00$ & - \\[2ex]
$\Upsilon(11020)$ & $6^{3}S_{1}$ & $BB$                       & $4.18$  & $5.28$   & - \\
                  &              & $BB^{\ast}$                & $15.49$ & $19.57$  & - \\
                  &              & $BB_{1}$                   & $40.08$ & $50.64$  & - \\
                  &              & $BB_{1}'$                  & $3.95$  & $4.98$   & - \\
                  &              & $B^{\ast}B^{\ast}$         & $11.87$ & $14.99$  & - \\
                  &              & $B_{s}B_{s}$               & $0.07$  & $0.09$   & - \\
                  &              & $B_{s}B_{s}^{\ast}$        & $1.50$  & $1.89$   & - \\
                  &              & $B_{s}^{\ast}B_{s}^{\ast}$ & $2.02$  & $2.56$   & - \\
$79\pm16$         &              & total                      & $79.16$ & $100.00$ & - \\
\hline
\hline
\end{tabular}
\caption {\label{tab:strongUps} Open $b$-flavored strong decay widths, in MeV, 
and branchings, in $\%$, of the $\Upsilon(4S)$, $\Upsilon(10860)$ and 
$\Upsilon(11020)$ states. Experimental data are taken from Ref.~\cite{PDG2014}.}
\end{center}
\end{table}

Table~\ref{tab:strongUps} shows the open-flavor strong decay widths of
the $\Upsilon(4S)$, $\Upsilon(10860)$ and $\Upsilon(11020)$ states. We calculate these
decays using a version of the $^{3}P_{0}$ model in which the strength $\gamma$
of the decay interaction scales as the reduced mass of the $q\bar{q}$-pair of
the decaying meson~\cite{Segovia:2012cd}. See Appendix~\ref{app:decays} for
further details. Following Eq.~(\ref{eq:fitgamma}), the value of $\gamma$ of the
$^{3}P_{0}$ model is $0.205$ in the bottomonium sector. One can see that the
general trend of the total decay widths is well reproduced.

The $\Upsilon(4S)$ is the first $1^{--}$ bottomonium state above the $B\bar{B}$
threshold, $10.56\,{\rm GeV}$. This state only decays into the $B\bar{B}$ final
channel. We have incorporated the isospin breaking via the experimental masses.
In Table~\ref{tab:strongUps} we compare the theoretical branching fractions with
the experimental ones for the two possible channels $B^{+}B^{-}$ and
$B^{0}\bar{B}^{0}$. Despite the mass of the $\Upsilon(4S)$ is very close to the 
thresholds, the difference between branching fractions of both channels is 
negligible due to the small difference between masses of the $B^{\pm}$ and 
$B^{0}$.

The possible two-body final decay channels of the $\Upsilon(10860)$ are $BB$, 
$BB^{\ast}$, $B^{\ast}B^{\ast}$, $B_{s}B_{s}$ $B_{s}B_{s}^{\ast}$
$B_{s}^{\ast}B_{s}^{\ast}$. The $^{3}P_{0}$ decay model predicts a total width
and branching fractions that are compatible with the experimental data
except for two cases in which the disagreement is quite strong. The decay 
channel $B^{\ast}B^{\ast}$ appears to be suppressed in the $^{3}P_{0}$ model 
whereas seems to be the dominant one attending to the experimental data. This 
is due to the small value of the overlap integral between the wave functions in 
our model. On the other hand, the theoretical branching fraction ${\cal 
B}(\Upsilon(10860)\to B_{s}^{(\ast)}B_{s}^{(\ast)})$ is roughly a factor $2$ 
bigger than the one measured experimentally. This is because we overestimate 
the rates of the $B_{s}B_{s}$, $B_{s}B_{s}^{\ast}$ and 
$B_{s}^{\ast}B_{s}^{\ast}$ final decay channels despite the order of magnitude 
is correctly given.

There is no experimental data about the open $b$-flavored strong decays of the
$\Upsilon(11020)$ resonance. Only its total decay width is known
experimentally~\cite{PDG2014} and the prediction of the $^{3}P_{0}$ model is in
very good agreement with such figure. The final decay channels $BB^{\ast}$,
$BB_{1}$ and $B^{\ast}B^{\ast}$ appear to be dominant in our model. The partial 
widths of the remaining decay channels are an order of magnitude smaller.

%%%%%%%%%%%%%%%%%%%%%%%%%%%%%%%%%%%%%%%%%%%%%%%%%%%%%%%%%%%%%%%%%%%%%%%%%%%%%%%%

\subsection{The $D$-wave levels}

Up to now, we have presented our theoretical results for the $S$- and $P$-wave 
bottomonium states. It is of some interest to go beyond this. For instance, a 
key test of the nonrelativistic potential description of bottomonium is the 
confirmation of the predicted $D$-wave levels. For this reason we perform in 
this section a theoretical study of the mesons $\eta_{b2}(^{1}D_{2})$, 
$\Upsilon(^{3}D_{1})$, $\Upsilon_{2}(^{3}D_{2})$ and $\Upsilon_{3}(^{3}D_{3})$.

Despite the $S$-wave and $P$-wave bottomonium states were first observed in the
1970s and 1980s, the triplet $\Upsilon(1^{3}D_{J})$ has been observed
recently~\cite{Bonvicini:2004yj} distinguishing only the $\Upsilon(1^{3}D_{2})$
state~\cite{delAmoSanchez:2010kz}. The mass of the $\Upsilon_{2}(1D)$ was 
measured to be $(10164.5\pm0.8\pm0.5)\,{\rm MeV}$. One can consider that the 
mass of the spin-triplet $\Upsilon(1^{3}D_{J})$ should be around this value 
assuming that the relativistic corrections are small in the bottomonium sector. 
Our theoretical mass for this spin-triplet is $10123\,{\rm MeV}$ which is lower 
than the experimental data and also with respect other theoretical
predictions~\cite{Eichten:1980mw,Gupta:1982kp,Gupta:1984jb,Gupta:1984um,
Gupta:1986xta,Moxhay:1983vu,Kwong:1988ae}.

It is inferred from Table~\ref{tab:predmassesbb} that the next set of
spin-triplet $D$-wave levels is expected in the range of $10419\,{\rm MeV}$,
and for the second radial excitation in the range of $10658\,{\rm MeV}$.
Contrary to the $1D$ multiplet, these two values are in better agreement with
those predicted by other quark models. The fine structure within the $D$-wave
multiplets is predicted to be somewhat smaller in the present model than in
those of other authors, but there is a general agreement about these
mass-splittings to be in the order of $\pm10\,{\rm MeV}$.

Despite the spin-singlet $D$-wave states, $\eta_{b2}$, are expected to lie very 
close in mass to the spin-weighted average of the spin-triplet $D$-wave states,
$\Upsilon_{J}$, they are still missing experimentally. One possibility to find 
the $1^{1}D_{2}$ and $2^{1}D_{2}$ states is studying the E1 radiative decays 
$h_{b}(2P)\to \gamma\eta_{b2}(1D)$ and $h_{b}(3P)\to \gamma\eta_{b2}(2D)$. Our 
prediction for both decays is in the order of $5\,{\rm keV}$ (see 
Table~\ref{tab:E1hb}). However, a better possibility is studying the decays 
$\eta_{b2}(1D)\to \gamma h_{b}(1P)$ and $\eta_{b2}(2D)\to \gamma h_{b}(2P)$ 
because the final states are well established in the PDG and our model predicts 
decay rates of few tens of keV (see Table~\ref{tab:E1etab2}). The decay rate of 
the $\eta_{b2}(2D)\to\gamma h_{b}(1P)$ transition is an order of magnitude 
smaller than the previous ones and thus its observation seems to be complicated.

\begin{table}[!t]
\begin{center}
\begin{tabular}{llrr}
\hline
\hline
\tstrut
Initial state & Final state & $\Gamma_{\rm The.}$ & ${\cal B}_{\rm The.}$ \\
& & (keV) & $(\times10^{-2})$ \\
\hline
\tstrut
$\eta_{b2}(1D)$ & $gg$                 & $0.37$  & $2.07$   \\   
                & $\gamma h_{b}(1P)$   & $17.23$ & $96.58$  \\
                & $\pi\pi\eta_{b}(1S)$ & $0.24$  & $1.35$   \\
                & total                & $17.84$ & $100.00$ \\[2ex]
$\eta_{b2}(2D)$ & $gg$                     & $0.67$              & $3.57$              \\
                & $\gamma h_{b}(1P)$       & $4.15$              & $22.13$             \\
                & $\gamma h_{b}(2P)$       & $11.66$             & $62.18$             \\
                & $\gamma h_{b3}(1F)$      & $2.20$              & $11.73$             \\
                & $\gamma\Upsilon_{2}(1D)$ & $1.27\times10^{-4}$ & $6.77\times10^{-4}$ \\
                & $\gamma\Upsilon_{3}(1D)$ & $5.30\times10^{-6}$ & $2.83\times10^{-5}$ \\
                & $\pi\pi\eta_{b}(1S)$     & $6.44\times10^{-2}$ & $0.34$              \\
                & $\pi\pi\eta_{b}(2S)$     & $8.39\times10^{-3}$ & $4.47\times10^{-2}$ \\
                & total                    & $18.75$             & $100.00$            \\
\hline
\hline
\end{tabular}
\caption{\label{tab:E1etab2} Decay widths and branching fractions 
of annihilation rates, radiative decays and hadronic transitions 
for the $\eta_{b2}(1D)$ and $\eta_{b2}(2D)$ states. There is
no experimental data available.}
\end{center}
\end{table}

\begin{table}[!t]
\begin{center}
\scalebox{0.85}{\begin{tabular}{llrrc}
\hline
\hline
\tstrut
Initial state & Final state & $\Gamma_{\rm The.}$ & ${\cal B}_{\rm The.}$ & ${\cal B}_{\rm Exp.}$~\cite{PDG2014} \\
& & (keV) & $(\times10^{-2})$ & $(\times10^{-2})$ \\
\hline
\tstrut
$\Upsilon(1D)$     & $e^{+}e^{-}$          & $1.40\times10^{-3}$ & $3.17\times10^{-3}$ & - \\    
                   & $3g$                  & $9.97$              & $22.57$             & - \\
                   & $\gamma\chi_{b0}(1P)$ & $20.98$             & $47.49$             & - \\
                   & $\gamma\chi_{b1}(1P)$ & $12.29$             & $27.82$             & - \\
                   & $\gamma\chi_{b2}(1P)$ & $0.65$              & $1.47$              & - \\
                   & $\pi\pi\Upsilon(1S)$  & $0.29$              & $0.66$              & $<0.82$~\cite{delAmoSanchez:2010kz} \\
                   & total                 & $44.18$             & $100.00$            & - \\[2ex]
$\Upsilon_{2}(1D)$ & $3g$                  & $0.62$  & $2.13$  & - \\
                   & $\gamma\chi_{b1}(1P)$ & $21.95$ & $75.46$ & - \\
                   & $\gamma\chi_{b2}(1P)$ & $6.23$  & $21.42$ & - \\
                   & $\pi\pi\Upsilon(1S)$  & $0.29$  & $1.00$  & $0.99^{+0.23}_{-0.21}\pm0.09$ \\
                   & total                 & $29.09$ & $100.0$ & - \\[2ex]
$\Upsilon_{3}(1D)$ & $3g$                  & $0.22$  & $0.87$   & - \\
                   & $\gamma\chi_{b2}(1P)$ & $24.74$ & $97.98$  & - \\
                   & $\pi\pi\Upsilon(1S)$  & $0.29$  & $1.15$   & $<0.62$~\cite{delAmoSanchez:2010kz} \\
                   & total                 & $25.25$ & $100.00$ & - \\[2ex]
$\Upsilon(2D)$     & $e^{+}e^{-}$          & $2.50\times10^{-3}$ & $8.24\times10^{-3}$ & - \\
                   & $3g$                  & $9.69$              & $31.93$             & - \\
                   & $\gamma\chi_{b0}(1P)$ & $3.52$              & $11.60$             & - \\
                   & $\gamma\chi_{b1}(1P)$ & $1.58$              & $5.21$              & - \\
                   & $\gamma\chi_{b2}(1P)$ & $6.08\times10^{-2}$ & $0.20$              & - \\
                   & $\gamma\chi_{b0}(2P)$ & $8.35$              & $27.52$             & - \\
                   & $\gamma\chi_{b1}(2P)$ & $4.84$              & $15.95$             & - \\
                   & $\gamma\chi_{b2}(2P)$ & $0.24$              & $0.79$              & - \\
                   & $\gamma\chi_{b2}(1F)$ & $2.05$              & $6.76$              & - \\
                   & $\gamma\eta_{b2}(1D)$ & $4.46\times10^{-6}$ & $1.47\times10^{-5}$ & - \\
                   & $\pi\pi\Upsilon(1S)$  & $7.10\times10^{-3}$ & $2.34\times10^{-2}$ & - \\
                   & $\pi\pi\Upsilon(2S)$  & $3.98\times10^{-3}$ & $1.31\times10^{-2}$ & - \\
                   & total                 & $30.34$             & $100.00$            & - \\[2ex]
$\Upsilon_{2}(2D)$ & $3g$                  & $0.61$              & $3.26$              & - \\
                   & $\gamma\chi_{b1}(1P)$ & $3.43$              & $18.35$             & - \\
                   & $\gamma\chi_{b2}(1P)$ & $0.80$              & $4.28$              & - \\
                   & $\gamma\chi_{b1}(2P)$ & $9.10$              & $48.69$             & - \\
                   & $\gamma\chi_{b2}(2P)$ & $2.55$              & $13.64$             & - \\
                   & $\gamma\chi_{b2}(1F)$ & $0.25$              & $1.34$              & - \\
                   & $\gamma\chi_{b3}(1F)$ & $1.93$              & $10.33$             & - \\
                   & $\gamma\eta_{b2}(1D)$ & $1.35\times10^{-4}$ & $7.22\times10^{-4}$ & - \\                   
                   & $\pi\pi\Upsilon(1S)$  & $1.54\times10^{-2}$ & $8.24\times10^{-2}$ & - \\
                   & $\pi\pi\Upsilon(2S)$  & $4.52\times10^{-3}$ & $2.42\times10^{-2}$ & - \\
                   & total                 & $18.69$             & $100.00$            & - \\[2ex]
$\Upsilon_{3}(2D)$ & $3g$                  & $1.25$              & $7.82$              & - \\
                   & $\gamma\chi_{b2}(1P)$ & $3.80$              & $23.79$             & - \\
                   & $\gamma\chi_{b2}(2P)$ & $10.70$             & $66.98$             & - \\
                   & $\gamma\chi_{b2}(1F)$ & $4.96\times10^{-3}$ & $3.10\times10^{-2}$ & - \\
                   & $\gamma\chi_{b3}(1F)$ & $0.19$              & $1.19$              & - \\
                   & $\gamma\eta_{b2}(1D)$ & $5.68\times10^{-4}$ & $3.56\times10^{-3}$ & - \\
                   & $\gamma\eta_{b2}(2D)$ & $8.73\times10^{-7}$ & $5.46\times10^{-6}$ & - \\
                   & $\pi\pi\Upsilon(1S)$  & $2.55\times10^{-2}$ & $0.16$              & - \\
                   & $\pi\pi\Upsilon(2S)$  & $5.13\times10^{-3}$ & $3.21\times10^{-2}$ & - \\
                   & total                 & $15.98$             & $100.00$            & - \\
\hline
\hline
\end{tabular}}
\caption{\label{tab:E1Dwaves} Decay widths and branching fractions 
of annihilation rates, radiative decays and hadronic transitions 
for the $1D$ and $2D$ states of the $\Upsilon$, $\Upsilon_{2}$ and 
$\Upsilon_{3}$ mesons. An estimate of the theoretical total
decay width is provided. Experimental data are taken from Ref.~\cite{PDG2014}.}
\end{center}
\end{table}

Table~\ref{tab:E1Dwaves} shows the decay widths and branching fractions of 
annihilation rates, radiative decays and hadronic transitions for the $1D$ and 
$2D$ states of the $\Upsilon$, $\Upsilon_{2}$ and $\Upsilon_{3}$ mesons. As one 
can see in the Table, the di-electron decay rates of the $\Upsilon$ $D$-wave 
states are two orders of magnitude smaller than those of the $S$-wave states 
(see Tables~\ref{tab:E1Supsilon} and~\ref{tab:E1Supsilon_v2}). The $\Upsilon$ 
family can be studied easily via $e^{+}e^{-}$ annihilation as they have the same 
quantum numbers of the emitted virtual photon. However, the production rate in 
this reaction is related with the leptonic width and we have seen that they are 
very small for the $\Upsilon$ $D$-wave states. This is the reason why there 
is no experimental confirmation of the $1^{--}$ $D$-wave states. At this point, 
it is also worthy to remind that the di-electron width of a $Q\bar{Q}$ meson is 
orders of magnitude larger than the corresponding one for a multi-quark 
system~\cite{Badalian:1985es}.

We have mentioned above that the constant $C_{2}$ of Eq.~(\ref{HofE1E1}) is 
fixed through the transition $\Upsilon_{2}(1D)\to\pi\pi\Upsilon(1S)$. This 
decay implies a $D\to S$ transition and thus it is the cleanest way to 
determine the constant $C_{2}$. Moreover, it is the only decay of this kind that 
presents a measurement of its branching fraction in the PDG~\cite{PDG2014}. As 
one can see in Table~\ref{tab:E1Dwaves}, the decay widths of the 
$\Upsilon(1D)\to\pi\pi\Upsilon(1S)$, $\Upsilon_{2}(1D)\to\pi\pi\Upsilon(1S)$ 
and $\Upsilon_{3}(1D)\to\pi\pi\Upsilon(1S)$ transitions are very similar, in 
the order of tenths of keV. The predicted branching fractions are in reasonable good
agreement with the experimental upper limits~\cite{delAmoSanchez:2010kz}.

One can inferred from Table~\ref{tab:E1Dwaves} that the radiative decay rates 
of the $1D$ states are dominant. The $\Upsilon(1D)$ state decays radiatively 
into the $\chi_{bJ}(1P)$ with $J=0,\,1,\,2$; the $\Upsilon_{2}(1D)$ only for 
$J=1,\,2$; and the $\Upsilon_{3}(1D)$ only for $J=2$. The partial widths of the 
$\Upsilon(1D)\to\gamma\chi_{b0}(1P)$, $\Upsilon_{2}(1D)\to\gamma\chi_{b1}(1P)$ 
and $\Upsilon_{3}(1D)\to\gamma\chi_{b2}(1P)$ processes are the largest ones 
with values around $20-25\,{\rm keV}$. An interesting feature can be deduced 
from here, the strength of the radiative decay into $\chi_{bJ_{f}}$ final meson 
depends on the total-spin $J_{i}$ of the initial $\Upsilon_{J_{i}}$ being of the 
same order of magnitude when $J_{i}=J_{f}+1$. This implies that one needs to 
design an experiment involving very high spin resonances in order to find 
simultaneously the $\Upsilon_{J}$ states in radiative decays. This would 
explain why the spin-triplet $1D$ multiplet has been observed for the first time 
with enough significance in the $\pi\pi\Upsilon(1S)$ final decay
channel~\cite{delAmoSanchez:2010kz}.

Table~\ref{tab:E1Dwaves} shows that the $2D$ states have similar decay features 
than the $1D$ states: i) they can decay into $\pi\pi\Upsilon(1S)$ and 
$\pi\pi\Upsilon(2S)$ final channels but with partial widths much smaller than 
those of the $1D$ states; ii) the radiative decays are the dominant ones but 
now the largest decay rate is in the order of $10\,{\rm keV}$; and iii) one can 
observe again that the strongest radiative decay into $\chi_{bJ_{f}}$ is the one 
in which the total-spin $J_{i}$ of the initial $\Upsilon_{J_{i}}$ is equal to 
$J_{f}+1$. The difference here is that this fact was observed for the $1D\to 
1P$ transitions and now it is fulfilled by the $2D\to 2P$ transitions.

It is also inferred from Table~\ref{tab:E1Dwaves} that the $1D$ and $2D$ states 
of the $\Upsilon$, $\Upsilon_{2}$ and $\Upsilon_{3}$ mesons are quite narrow 
with total decay widths in the order of $15-45\,{\rm keV}$. Moreover, the decay 
rate of their annihilation into gluons is not relevant except for the $1D$ and 
$2D$ $\Upsilon$ states with branching fractions of $23\%$ and $32\%$, 
respectively.

\begin{table}[!t]
\begin{center}
\begin{tabular}{ccccrrrr}
\hline
\hline
\multicolumn{4}{c}{Decay chain} & ${\cal B}_{1}$ & ${\cal B}_{2}$ & ${\cal
B}_{3}$ & ${\cal B}_{\rm Tot.}$ \\
& & & & $(\%)$ & $(\%)$ & $(\%)$ & $(10^{-6})$ \\
\hline
\tstrut
$3^{3}S_1$ & $\to 2^{3}P_{0}$ & $\to 2^{3}S_{1}$ & $\to 1^{3}P_{0}$ & $5.96$  & $0.54$  & $3.41$  & $10.98$ \\
           &                  &                  & $\to 1^{3}P_{1}$ & $5.96$  & $0.54$  & $5.75$  & $18.51$ \\
           &                  &                  & $\to 1^{3}P_{2}$ & $5.96$  & $0.54$  & $6.50$  & $20.92$ \\
           &                  & $\to 1^{3}D_{1}$ & $\to 1^{3}P_{0}$ & $5.96$  & $0.031$ & $47.49$ & $8.75$ \\
           &                  &                  & $\to 1^{3}P_{1}$ & $5.96$  & $0.031$ & $27.82$ & $5.12$ \\
           &                  &                  & $\to 1^{3}P_{2}$ & $5.96$  & $0.031$ & $1.47$  & $0.27$ \\[2ex]
$3^{3}S_1$ & $\to 2^{3}P_{1}$ & $\to 2^{3}S_{1}$ & $\to 1^{3}P_{0}$ & $10.48$ & $11.91$ & $3.41$  & $425.63$ \\
           &                  &                  & $\to 1^{3}P_{1}$ & $10.48$ & $11.91$ & $5.75$  & $717.70$ \\
           &                  &                  & $\to 1^{3}P_{2}$ & $10.48$ & $11.91$ & $6.50$  & $811.31$ \\
           &                  & $\to 1^{3}D_{1}$ & $\to 1^{3}P_{0}$ & $10.48$ & $0.31$  & $47.49$ & $154.29$ \\
           &                  &                  & $\to 1^{3}P_{1}$ & $10.48$ & $0.31$  & $27.82$ & $90.38$ \\
           &                  &                  & $\to 1^{3}P_{2}$ & $10.48$ & $0.31$  & $1.47$  & $4.78$ \\
           &                  & $\to 1^{3}D_{2}$ & $\to 1^{3}P_{1}$ & $10.48$ & $0.95$  & $75.46$ & $751.28$ \\
           &                  &                  & $\to 1^{3}P_{2}$ & $10.48$ & $0.95$  & $21.42$ & $213.26$ \\[2ex]
$3^{3}S_1$ & $\to 2^{3}P_{2}$ & $\to 2^{3}S_{1}$ & $\to 1^{3}P_{0}$ & $12.60$ & $12.86$ & $3.41$  & $552.54$ \\
           &                  &                  & $\to 1^{3}P_{1}$ & $12.60$ & $12.86$ & $5.75$  & $931.71$ \\
           &                  &                  & $\to 1^{3}P_{2}$ & $12.60$ & $12.86$ & $6.50$  & $1053.23$ \\
           &                  & $\to 1^{3}D_{1}$ & $\to 1^{3}P_{0}$ & $12.60$ & $0.015$ & $47.49$ & $9.22$ \\
           &                  &                  & $\to 1^{3}P_{1}$ & $12.60$ & $0.015$ & $27.82$ & $5.40$ \\
           &                  &                  & $\to 1^{3}P_{2}$ & $12.60$ & $0.015$ & $1.47$  & $0.29$ \\
           &                  & $\to 1^{3}D_{2}$ & $\to 1^{3}P_{1}$ & $12.60$ & $0.26$  & $75.46$ & $247.21$ \\
           &                  &                  & $\to 1^{3}P_{2}$ & $12.60$ & $0.26$  & $21.42$ & $70.17$ \\
           &                  & $\to 1^{3}D_{3}$ & $\to 1^{3}P_{2}$ & $12.60$ & $1.51$  & $97.98$ & $1864.17$ \\
\hline
\hline
\end{tabular}
\caption{\label{tab:decaychainDwave_v1} Radiative decay chains involving the
photon cascades $3S\to2P\to2S\to1P$ and $3S\to2P\to1D\to1P$. The branching
fractions are ${\cal B}_{1}={\cal B}(3^{3}S_{1}\to 2^{3}P_{J}+\gamma)$, ${\cal
B}_{2}={\cal B}(2^{3}P_{J}\to 2^{3}S_{1}+\gamma)$ or $={\cal B}(2^{3}P_{J}\to
1^{3}D_{J}+\gamma)$, ${\cal B}_{3}={\cal B}(2^{3}S_{1}\to 1^{3}P_{J}+\gamma)$ or
$={\cal B}(1^{3}D_{J}\to 1^{3}P_{J}+\gamma)$, and ${\cal B}_{\rm Tot.}={\cal
B}_{1}\times{\cal B}_{2}\times{\cal B}_{3}$.}
\end{center}
\end{table}

Photon cascade processes are usually used in order to study conventional
bottomonium states which are located below the open $b$-flavored threshold. If
we focus on the photon cascades starting from the $\Upsilon(3S)$, the usual
process is $3S\to2P\to2S$ which is experimentally identified via the subsequent
$e^{+}e^{-}$ or $\mu^{+}\mu^{-}$ decay of the $2S$ state. However, if the
three-photon cascade $3S\to2P\to2S\to1P$ can be observed, there is hope for
observing the corresponding $3S\to2P\to1D\to1P$ and thus a new possibility of
studying the $1D$-multiplet appears. It is worth to mention here that all the
radiative decays corresponding to the $3S\to2P\to2S\to1P$ decay chain have been
measured separately. The combined branching fractions of the three-photon
cascades $3S\to2P\to2S\to1P$ and $3S\to2P\to1D\to1P$ are shown in
Table~\ref{tab:decaychainDwave_v1}. The most prominent cascades involving $1D$
states are:
\begin{equation}
\begin{split}
&
3^{3}S_{1}\to2^{3}P_{1}\to1^{3}D_{2}\to1^{3}P_{1} \quad ({\cal
B}=7.51\times10^{-4}), \\
&
3^{3}S_{1}\to2^{3}P_{2}\to1^{3}D_{3}\to1^{3}P_{2} \quad ({\cal
B}=18.64\times10^{-4}),
\end{split}
\end{equation}
followed by
\begin{equation}
\begin{split}
&
3^{3}S_{1}\to2^{3}P_{1}\to1^{3}D_{1}\to1^{3}P_{0} \quad ({\cal
B}=1.54\times10^{-4}), \\
&
3^{3}S_{1}\to2^{3}P_{1}\to1^{3}D_{2}\to1^{3}P_{2} \quad ({\cal
B}=2.13\times10^{-4}), \\
&
3^{3}S_{1}\to2^{3}P_{2}\to1^{3}D_{2}\to1^{3}P_{1} \quad ({\cal
B}=2.47\times10^{-4}),
\end{split}
\end{equation}
and with
\begin{equation}
\begin{split}
&
3^{3}S_{1}\to2^{3}P_{1}\to1^{3}D_{1}\to1^{3}P_{1} \quad ({\cal
B}=0.90\times10^{-4}), \\
&
3^{3}S_{1}\to2^{3}P_{2}\to1^{3}D_{2}\to1^{3}P_{2} \quad ({\cal
B}=0.70\times10^{-4}),
\end{split}
\end{equation}
also significant. One can conclude that the range of possibilities is large
enough in order to disentangle the masses of the $1D$ spin-triplet members in
the near future. The $1^{3}D_{2}$ and $1^{3}D_{3}$ states have more chances to
be observed than the $1^{3}D_{1}$ in this kind of decay chains. The branching 
fractions associated with the 
$3^{3}S_{1}\to2^{3}P_{1}\to1^{3}D_{2}\to1^{3}P_{1}$ and 
$3^{3}S_{1}\to2^{3}P_{2}\to1^{3}D_{3}\to1^{3}P_{2}$ photon cascade processes 
that involve, respectively, the $1^{3}D_{2}$ and $1^{3}D_{3}$ states are $5$ 
and $12$ times larger than the most important three-photon cascade involving the
$1^{3}D_{1}$ state, $3^{3}S_{1}\to2^{3}P_{1}\to1^{3}D_{1}\to1^{3}P_{0}$.

\begin{table}[!t]
\begin{center}
\begin{tabular}{cccrrr}
\hline
\hline
\multicolumn{3}{c}{Decay chain} & ${\cal B}_{1}$ & ${\cal B}_{2}$ & ${\cal 
B}_{\rm Tot.}$ \\
& & & $(\%)$ & $(\%)$ & $(10^{-4})$ \\
\hline
\tstrut
$3^{3}P_{0}$ & $\to 2^{3}D_{1}$ & $\to 2^{3}P_{0}$ & $0.14$ & $27.52$ & $3.85$ \\
             &                  & $\to 2^{3}P_{1}$ & $0.14$ & $15.95$ & $2.23$ \\
             &                  & $\to 2^{3}P_{2}$ & $0.14$ & $0.79$  & $0.11$ \\[2ex]
$3^{3}P_{1}$ & $\to 2^{3}D_{1}$ & $\to 2^{3}P_{0}$ & $0.85$ & $27.52$ & $23.39$ \\
             &                  & $\to 2^{3}P_{1}$ & $0.85$ & $15.95$ & $13.56$ \\
             &                  & $\to 2^{3}P_{2}$ & $0.85$ & $0.79$  & $0.67$ \\
             & $\to 2^{3}D_{2}$ & $\to 2^{3}P_{1}$ & $2.24$ & $48.69$ & $109.07$ \\
             &                  & $\to 2^{3}P_{2}$ & $2.24$ & $13.64$ & $30.55$ \\[2ex]
$3^{3}P_{2}$ & $\to 2^{3}D_{1}$ & $\to 2^{3}P_{0}$ & $0.13$ & $27.52$ & $3.58$ \\
             &                  & $\to 2^{3}P_{1}$ & $0.13$ & $15.95$ & $2.07$ \\
             &                  & $\to 2^{3}P_{2}$ & $0.13$ & $0.79$  & $0.10$ \\
             & $\to 2^{3}D_{2}$ & $\to 2^{3}P_{1}$ & $0.56$ & $48.69$ & $27.27$ \\
             &                  & $\to 2^{3}P_{2}$ & $0.56$ & $13.64$ & $7.64$ \\
             & $\to 2^{3}D_{3}$ & $\to 2^{3}P_{2}$ & $2.97$ & $66.98$ & $198.93$ \\
\hline
\hline
\end{tabular}
\caption{\label{tab:decaychainDwave_v2} Radiative decay chains involving the
photon cascades $3P\to2D\to2P$. The branching fractions are ${\cal B}_{1}={\cal
B}(3^{3}P_{J}\to 2^{3}D_{J'}+\gamma)$, ${\cal B}_{2}={\cal B}(2^{3}D_{J'}\to
2^{3}P_{J''}+\gamma)$, and ${\cal B}_{\rm Tot.}={\cal B}_{1}\times{\cal
B}_{2}$.}
\end{center}
\end{table}

A similar game can be played in order to give some insight on the most
plausible photon cascades to study the spin-triplet $2D$ states.
Table~\ref{tab:decaychainDwave_v2} shows the two-photon cascades starting from
the $\chi_{bJ}(3P)$. As we have mentioned above, a new structure centered at a
mass of $10.5\,{\rm GeV}$ has been interpreted as the $\chi_{bJ}(3P)$
system~\cite{Aad:2011ih,aba:2012}. This structure is still below open 
$b$-flavored threshold and it should naturally decay into $2D$ states by E1 
radiative transitions, the second photon comes from the radiative decays of the 
$2D$ states into the well established $\chi_{bJ}(2P)$ mesons. As one can see in
Table~\ref{tab:decaychainDwave_v2}, the most prominent two-photon cascades are:
\begin{equation}
\begin{split}
&
3^{3}P_{1}\to 2^{3}D_{2}\to 2^{3}P_{1} \quad ({\cal B}=10.91\times10^{-3}), \\
&
3^{3}P_{2}\to 2^{3}D_{3}\to 2^{3}P_{2} \quad ({\cal B}=19.89\times10^{-3}),
\end{split}
\end{equation}
followed by
\begin{equation}
\begin{split}
&
3^{3}P_{1}\to 2^{3}D_{2}\to 2^{3}P_{2} \quad ({\cal B}=3.06\times10^{-3}), \\
&
3^{3}P_{2}\to 2^{3}D_{2}\to 2^{3}P_{1} \quad ({\cal B}=2.73\times10^{-3}).
\end{split}
\end{equation}
The two-photon cascades involving the $2^{3}D_{1}$ state present branching
fractions smaller than the ones shown above. The two most important decay
chains involving this state are
\begin{equation}
\begin{split}
&
3^{3}P_{1}\to 2^{3}D_{1}\to 2^{3}P_{0} \quad ({\cal B}=2.34\times10^{-3}), \\
&
3^{3}P_{1}\to 2^{3}D_{1}\to 2^{3}P_{1} \quad ({\cal B}=1.36\times10^{-3}).
\end{split}
\end{equation}

%%%%%%%%%%%%%%%%%%%%%%%%%%%%%%%%%%%%%%%%%%%%%%%%%%%%%%%%%%%%%%%%%%%%%%%%%%%%%%%%
%%%%%%%%%%%%%%%%%%%%%%%%%%%%%%%%%%%%%%%%%%%%%%%%%%%%%%%%%%%%%%%%%%%%%%%%%%%%%%%%

\section{SUMMARY AND CONCLUSIONS}
\label{sec:conclusions}

We have revisited the bottomonium spectrum motivated by the experimental 
progress in the last few years on determining new conventional and 
unconventional states in this sector. Our approach is a nonrelativistic 
constituent quark model whose model parameters are constrained by other quark 
sectors, from light to heavy, and thus our description of the bottomonium is, 
in this sense, parameter-free.

The bottomonium spectrum predicted by our quark model is in a global agreement
with the experimental data. Moreover, we have provided a large number of 
electromagnetic, strong and hadronic decays showing that our results are in 
reasonable agreement with the available experimental data in most of the cases. 
Amongst the results we describe, the following are of particular interest.

Our value for the mass of the $\eta_{b}(2S)$ is within the mass range given by 
the BaBar Collaboration and slightly lower than the CLEO estimation. The 
hyperfine mass splitting between the singlet and triplet $2S$ states is 
consistent with the experimental data and also with lattice QCD computations. A 
prediction of the $\eta_{b}(3S)$ and its corresponding hyperfine mass splitting 
is provided. In order to give more insights about the better way to determine 
their properties experimentally, we have computed the decay widths and 
branching fractions of annihilation rates, radiative decays and hadronic 
transitions for the $\eta_{b}(1S)$, $\eta_{b}(2S)$ and $\eta_{b}(3S)$ states.

The masses predicted by our theoretical model for the $h_{b}$ states are located
at the spin-weighted average of the corresponding triplet $\chi_{bJ}$ states. 
This indicates that our hyperfine interaction is compatible with zero as should 
be from experimental observations. The decay widths of annihilation rates, 
radiative decays and hadronic transitions for the $h_{b}$ mesons have been also 
provided indicating that these states are narrow mesons with total decay widths 
of about $100\,{\rm keV}$.

The ATLAS and D0 Collaborations have reported very recently the average mass of 
the $\chi_{bJ}(3P)$ multiplet. They were not able to distinguish the different 
members of this multiplet, only the LHCb Collaboration has provided a mass 
estimation for the $\chi_{b1}(3P)$ state which is in very good agreement with 
our quark model result. We predict intra-multiplet splittings in the order of
$\sim\!\!10\,{\rm MeV}$. We have calculated decay widths of hadronic, radiative 
and annihilation into gluons processes concerning the $\chi_{bJ}(1P,2P,3P)$. In 
general, our theoretical results are in agreement with the available 
experimental data. Special attention deserves the predicted decay properties of 
the $\chi_{bJ}(3P)$ states which can help experimentalists to determine the 
properties of the different members of the multiplet. It is possible that we are 
overestimating the annihilation rates into gluons for the $\chi_{b0}(nP)$ states 
but we have not been able to give a definitive statement.

Our theoretical description of the $\Upsilon$ family has been exhaustive. We
have provided masses and also a wide range of decay properties. Focusing on the 
radiative transitions, we have computed the branching fractions for all 
experimentally available decays and combine them in a way that allows us to 
compare with the most updated experimental study. In general, our theoretical 
results agree with experiment although there are some cases in which the 
discrepancies are important. Focusing on the open-flavor strong decays, we 
achieve a global description of the partial and total decay widths with a 
version of the $^{3}P_{0}$ model in which the strength $\gamma$ is 
scale-dependent as a function of the reduced mass of the quark-antiquark pair 
of the decaying meson. Since the $\Upsilon(4S)$, $\Upsilon(10860)$ and 
$\Upsilon(11020)$ are above the $B\bar{B}$ threshold, the study of these states 
has to be done using more sophisticated approaches that incorporate the 
effect of meson-meson thresholds. This computation is beyond the scope of this 
work and thus we stress that the results herein for the above states have 
to be taken with care and leave the coupled-channels study for a future work.

We have investigated properties of the $D$-wave bottomonium levels and have made
suggestions for their observation. Up to now, there is only experimental
confirmation of the $1^{3}D_{2}$ state with a mass of about $10.16\,{\rm GeV}$.
According to our model, the next sets of the spin-triplet $D$-wave levels are
expected in the range of $10.42$ and $10.66\,{\rm GeV}$. The mass splittings
between members of the same multiplet are lower than $10\,{\rm MeV}$. The 
$^{1}D_{2}$ states can decay into the relatively new observed $h_{b}(1P)$ and
$h_{b}(2P)$ states via radiative transitions. The best prospects for studying
the $1^{3}D_{J}$ and $2^{3}D_{J}$ states appear to be their production via
multiphoton cascades. It is worthy to emphasize here that the strength of the 
radiative decay into $\chi_{bJ_{f}}$ final meson depends on the total-spin 
$J_{i}$ of the initial $\Upsilon_{J_{i}}$ being the strongest one the case in 
which $J_{i}=J_{f}+1$.

%%%%%%%%%%%%%%%%%%%%%%%%%%%%%%%%%%%%%%%%%%%%%%%%%%%%%%%%%%%%%%%%%%%%%%%%%%%%%%%%
%%%%%%%%%%%%%%%%%%%%%%%%%%%%%%%%%%%%%%%%%%%%%%%%%%%%%%%%%%%%%%%%%%%%%%%%%%%%%%%%

\begin{acknowledgments}
J. Segovia would like to thank Nora Brambilla and Antonio Vairo for insightful 
comments.
This work has been partially funded by Ministerio de Ciencia y Tecnolog\'\i a 
under Contract no. FPA2013-47443-C2-2-P, by the Spanish Excellence Network on 
Hadronic Physics FIS2014-57026-REDT and by the Spanish Ingenio-Consolider 2010
Program CPAN (CSD2007-00042).
J. Segovia acknowledges financial support from a postdoctoral IUFFyM contract 
of the Universidad de Salamanca, Spain; and from the Alexander von Humboldt 
Foundation.
P.G. Ortega acknowledges the financial support from the European Union's Marie 
Curie COFUND grant (PCOFUND-GA-2011-291783).
\end{acknowledgments}

%%%%%%%%%%%%%%%%%%%%%%%%%%%%%%%%%%%%%%%%%%%%%%%%%%%%%%%%%%%%%%%%%%%%%%%%%%%%%%%%
%%%%%%%%%%%%%%%%%%%%%%%%%%%%%%%%%%%%%%%%%%%%%%%%%%%%%%%%%%%%%%%%%%%%%%%%%%%%%%%%

\appendix

\section{CONSTITUENT QUARK MODEL}
\label{app:CQM}

We work within the framework of a constituent quark model proposed in
Ref.~\cite{Vijande:2004he} (see references~\cite{Valcarce:2005em} 
and~\cite{Segovia:2013wma} for reviews). This model describes quite well hadron 
phenomenology and hadronic 
reactions~\cite{Fernandez:1992xs,Garcilazo:2001md,Vijande:2004at}. Furthermore, 
it has been recently applied to mesons containing heavy quarks with great 
success, describing a wide range of physical observables which concern 
spectrum~\cite{Segovia:2008zz,Segovia:2009zz,Segovia:2015dia}, strong 
reactions~\cite{Segovia:2011zza,Segovia:2012cd,Segovia:2013kg} and weak 
decays~\cite{Segovia:2011dg,Segovia:2012yh,Segovia:2013sxa}.

We have mentioned above that in the heavy quark sector chiral symmetry is
explicitly broken and Goldstone-boson exchanges do not appear. Thus, one-gluon
exchange and confinement are the only interactions remaining. The one-gluon
exchange potential contains central, tensor and spin-orbit contributions given
by
\begin{widetext}
\begin{equation}
\begin{split}
&
V_{\rm OGE}^{\rm C}(\vec{r}_{ij}) =
\frac{1}{4}\alpha_{s}(\vec{\lambda}_{i}^{c}\cdot
\vec{\lambda}_{j}^{c})\left[ \frac{1}{r_{ij}}-\frac{1}{6m_{i}m_{j}} 
(\vec{\sigma}_{i}\cdot\vec{\sigma}_{j}) 
\frac{e^{-r_{ij}/r_{0}(\mu)}}{r_{ij}r_{0}^{2}(\mu)}\right], \\
& 
V_{\rm OGE}^{\rm T}(\vec{r}_{ij})=-\frac{1}{16}\frac{\alpha_{s}}{m_{i}m_{j}}
(\vec{\lambda}_{i}^{c}\cdot\vec{\lambda}_{j}^{c})\left[ 
\frac{1}{r_{ij}^{3}}-\frac{e^{-r_{ij}/r_{g}(\mu)}}{r_{ij}}\left( 
\frac{1}{r_{ij}^{2}}+\frac{1}{3r_{g}^{2}(\mu)}+\frac{1}{r_{ij}r_{g}(\mu)}\right)
\right]S_{ij}, \\
&
\begin{split}
V_{\rm OGE}^{\rm SO}(\vec{r}_{ij})= &  
-\frac{1}{16}\frac{\alpha_{s}}{m_{i}^{2}m_{j}^{2}}(\vec{\lambda}_{i}^{c} \cdot
\vec{\lambda}_{j}^{c})\left[\frac{1}{r_{ij}^{3}}-\frac{e^{-r_{ij}/r_{g}(\mu)}}
{r_{ij}^{3}} \left(1+\frac{r_{ij}}{r_{g}(\mu)}\right)\right] \times \\ & \times 
\left[((m_{i}+m_{j})^{2}+2m_{i}m_{j})(\vec{S}_{+}\cdot\vec{L})+
(m_{j}^{2}-m_{i}^{2}) (\vec{S}_{-}\cdot\vec{L}) \right],
\end{split}
\end{split}
\label{eq:OGEpot}
\end{equation}
\end{widetext}
where $r_{0}(\mu)=\hat{r}_{0}\frac{\mu_{nn}}{\mu_{ij}}$ and
$r_{g}(\mu)=\hat{r}_{g}\frac{\mu_{nn}}{\mu_{ij}}$ are regulators which depend on
$\mu_{ij}$, the reduced mass of the $q\bar{q}$ pair. The contact term of the
central potential has been regularized as
\begin{equation}
\delta(\vec{r}_{ij})\sim\frac{1}{4\pi 
r_{0}^{2}}\frac{e^{-r_{ij}/r_{0}}}{r_{ij}}.
\label{eq:delta}
\end{equation}

The wide energy range needed to provide a consistent description of light,
strange and heavy mesons requires an effective scale-dependent strong coupling
constant. We use the frozen coupling constant~\cite{Vijande:2004he}
\begin{equation}
\alpha_{s}(\mu)=\frac{\alpha_{0}}{\ln\left(
\frac{\mu^{2}+\mu_{0}^{2}}{\Lambda_{0 }^{2}} \right)},
\end{equation}
in which $\mu$ is the reduced mass of the $q\bar{q}$ pair and $\alpha_{0}$,
$\mu_{0}$ and $\Lambda_{0}$ are parameters of the model determined by a global
fit to the meson spectra.

The different pieces of the confinement potential are
\begin{equation}
\begin{split}
&
V_{\rm CON}^{\rm C}(\vec{r}_{ij})=\left[-a_{c}(1-e^{-\mu_{c}r_{ij}})+\Delta
\right] (\vec{\lambda}_{i}^{c}\cdot\vec{\lambda}_{j}^{c}), \\
&
\begin{split}
&
V_{\rm CON}^{\rm SO}(\vec{r}_{ij}) =
-(\vec{\lambda}_{i}^{c}\cdot\vec{\lambda}_{j}^{c}) \frac{a_{c}\mu_{c}e^{-\mu_{c}
r_{ij}}}{4m_{i}^{2}m_{j}^{2}r_{ij}} \times \\
&
\times 
\left[((m_{i}^{2}+m_{j}^{2})(1-2a_{s}) + 4m_{i}m_{j}(1-a_{s}))(\vec{S}_{+} 
\cdot\vec{L}) \right. \\
&
\left. \quad\,\, +(m_{j}^{2}-m_{i}^{2}) (1-2a_{s}) (\vec{S}_{-}\cdot\vec{L})
\right],
\end{split}
\end{split}
\end{equation}
where $a_{s}$ controls the mixture between the scalar and vector Lorentz
structures of the confinement. At short distances this potential presents a
linear behavior with an effective confinement strength
$\sigma=-a_{c}\,\mu_{c}\,(\vec{\lambda}^{c}_{i}\cdot \vec{\lambda}^{c}_{j})$,
while it becomes constant at large distances. This type of potential shows a
threshold defined by
\begin{equation}
V_{\rm thr}=\{-a_{c}+ \Delta\}(\vec{\lambda}^{c}_{i}\cdot
\vec{\lambda}^{c}_{j}).
\end{equation}
No $q\bar{q}$ bound states can be found for energies higher than this threshold.
The system suffers a transition from a color string configuration between two
static color sources into a pair of static mesons due to the breaking of the
color string and the most favored decay into hadrons.

Among the different methods to solve the Schr\"odinger equation in order to 
find the quark-antiquark bound states, we use the Gaussian Expansion
Method (GEM)~\cite{Hiyama:2003cu} which provides enough accuracy and simplifies 
the subsequent evaluation of the decay amplitude matrix elements.

This procedure provides the radial wave function solution of the Schr\"odinger
equation as an expansion in terms of basis functions
\begin{equation}
R_{\alpha}(r)=\sum_{n=1}^{n_{max}} c_{n}^\alpha \phi^G_{nl}(r),
\end{equation} 
where $\alpha$ refers to the channel quantum numbers. The coefficients,
$c_{n}^\alpha$, and the eigenvalue, $E$, are determined from the Rayleigh-Ritz
variational principle
\begin{equation}
\sum_{n=1}^{n_{max}} \left[\left(T_{n'n}^\alpha-EN_{n'n}^\alpha\right)
c_{n}^\alpha+\sum_{\alpha'}
\ V_{n'n}^{\alpha\alpha'}c_{n}^{\alpha'}=0\right],
\end{equation}
where $T_{n'n}^\alpha$, $N_{n'n}^\alpha$ and $V_{n'n}^{\alpha\alpha'}$ are the 
matrix elements of the kinetic energy, the normalization and the potential, 
respectively. $T_{n'n}^\alpha$ and $N_{n'n}^\alpha$ are diagonal, whereas the
mixing between different channels is given by
$V_{n'n}^{\alpha\alpha'}$.

Following Ref.~\cite{Hiyama:2003cu}, we employ Gaussian trial functions with
ranges  in geometric progression. This enables the optimization of ranges
employing a small number of free parameters. Moreover, the geometric
progression is dense at short distances, so that it enables the description of
the dynamics mediated by short range potentials. The fast damping of the
Gaussian tail does not represent an issue, since we can choose the maximal
range much longer than the hadronic size.

%%%%%%%%%%%%%%%%%%%%%%%%%%%%%%%%%%%%%%%%%%%%%%%%%%%%%%%%%%%%%%%%%%%%%%%%%%%%%%%%
%%%%%%%%%%%%%%%%%%%%%%%%%%%%%%%%%%%%%%%%%%%%%%%%%%%%%%%%%%%%%%%%%%%%%%%%%%%%%%%%

\section{DECAYS AND REACTIONS}
\label{app:decays}

\subsection{Radiative decays}
\label{subsec:radiative}

Electromagnetic E1 and M1 dominant multipole transitions have been studied
since the early days of hadron spectroscopy because they allow to access heavy
quarkonium states which are below open-flavor threshold. Moreover, they are
interesting by themselves because is an important tool to determine the internal
charge structure of hadrons and its quantum numbers. From a theoretical point 
of view, electromagnetic transitions have been treated traditionally within the 
potential model approach. However, in the last decade, progress has been made 
using effective field theories~\cite{Brambilla:2005zw, Brambilla:2012be}. The 
decay rate for E1 transitions between an initial state $n^{2S+1}L_J$ and a 
final state $n'\,^{2S'+1}L'_{J'}$ can be written as~\cite{Brambilla:2004jw}
\begin{equation}
\begin{split}
\Gamma_{E1} (n^{2S+1}&L_J  \rightarrow n'\,^{2S'+1}L'_{J'}) = \\
&= \frac{4\alpha e^2_bk^3}{3} (2J'+1) S_{fi}^E\,\delta_{SS'}\,|\mathcal
E_{fi}|^2\frac{E_f}{M_i},
\end{split}
\end{equation}
where $k$ is the emitted photon momentum, $E_{f}/M_{i}$ is a relativistic
correction with $M_i$ the mass of the initial state and $E_f$ the
energy of the final state. The statistical factor, $S_{fi}^E$, is given by 
\begin{equation}
S_{fi}^E={\rm max}(L,L')\left\{\begin{matrix} J & 1 & J' \\ L'& S & L
\end{matrix}\right\}^2.
\end{equation}
If the full momentum dependence is retained, the overlap integral, $\mathcal
E_{fi}$, is
\begin {equation}
{\cal E}_{fi} = \frac{3}{k}\int_{0}^{\infty} R_{\alpha'}(r)  
\left[ \frac{kr}{2}j_0\left(\frac{kr}{2}\right)-j_1\left(\frac{kr}{2}\right)
\right] R_{\alpha}(r) \, r^2 \,dr,
\label{eq:EfiE1}
\end{equation}
where $j_{i}(x)$ are the spherical Bessel functions of the first kind and 
$\alpha$ ($\alpha'$) are the initial (final) meson quantum numbers.

The M1 radiative transitions can be evaluated with the following expression
\begin{equation}
\begin{split}
\Gamma_{M1}(n^{2S+1}L_J & \rightarrow n'\,^{2S'+1}L'_{J'}) = \\
&= \frac{4\alpha e_b^2k^3}{3m_b^2} (2J'+1)S_{fi}^M |{\mathcal M_{fi}}|^2
\frac{E_f}{M_i},
\end{split}
\end{equation}
where we use the same notation as in the E1 transitions but now 
\begin{equation}
S_{fi}^M=6(2S+1)(2S'+1)
\left\{\begin{matrix}J & 1 & J' \\ S'& L & S \end{matrix}\right\}^2
\left\{\begin{matrix}1 & 1/2 & 1/2 \\ 1/2& S' & S \end{matrix}\right\}^2,
\end{equation}
and
\begin {eqnarray}
{\cal M}_{fi} = \int_{0}^{\infty} R_{\alpha'}(r) j_{0}\left(\frac{kr}{2}\right)
R_{\alpha}(r) \, r^2 \,dr.
\end{eqnarray}

\subsection{Annihilation decays}

The knowledge of annihilation decay rates is important for several reasons. 
First, this kind of decays allows to test the wave function at very short 
range. Second, the annihilation decays into gluons and light quarks make 
significant contributions to the total decay widths of some bottomonium states. 
Third, the annihilation decays into leptons or photons can be useful for 
the production and identification of resonances. And fourth, leptonic decay 
rates can help to distinguish between conventional mesons and multiquark 
structures which have much smaller di-electron widths~\cite{Badalian:1985es}.

The dominant contribution to the decay of quarkonium states into lepton pairs
proceed via a single virtual photon, as long as the mass of the initial meson
state is sufficiently small that the contribution of a virtual $Z$ can be
ignored. The leptonic width of $^{3}S_{1}$ bottomonium including radiative QCD
corrections is given by~\cite{Barbieri:1975ki}
\begin{equation}
\Gamma\left(n^{3}S_{1}\to e^{+}e^{-} \right) = \frac{4\alpha^{2}e^{2}_{b}
|R_{n}(0)|^2}{M^2_n} \left(1-\frac{16\alpha_{s}}{3\pi}\right),
\end{equation}
where $\alpha\simeq1/137$ is the fine-structure constant and $e_{b}=-1/3$ is
the charge of the bottom quark in units of the electron's charge. Similarly for
$D$-wave $1^{--}$ bottomonium states, the leading order decay width into 
$e^{+}e^{-}$ is given by~\cite{Novikov:1977dq} 
\begin{equation}
\Gamma\left( n^{3}D_{1}\to e^{+}e^{-} \right) =
\frac{25\alpha^{2}e_{b}^2}{2m_{b}^4M_{n}^{2}} |R_{n}^{''}(0)|^{2}.
\end {equation}
The leading QCD correction to this expression has been calculated in 
Ref.~\cite{Bradley:1980eh}, but we do not considered it here.

The annihilation decay rates into gluons and/or photons of the $^{3}S_{1}$ 
bottomonium states including radiative QCD corrections are given 
by~\cite{Kwong:1987ak,Kwong:1988ae}
\begin{equation}
\begin{split}
\Gamma(n^{3}S_{1}\to 3g) &= \frac{10(\pi^{2}-9)\alpha_{s}^{3}}{81\pi m_{b}^{2}}
\, |R_{nS}(0)|^{2} \left( 1-\frac{4.9\alpha_{s}}{\pi}\right), \\
\Gamma(n^{3}S_{1}\to \gamma gg) &= \frac{8(\pi^{2}-9)e_{b}^{2} \alpha 
\alpha_{s}^{2}}{9\pi m_{b}^{2}} \, |R_{nS}(0)|^{2}
\left( 1-\frac{7.4\alpha_{s}}{\pi}\right), \\
\Gamma(n^{3}S_{1}\to 3\gamma) &= \frac{4(\pi^{2}-9)e_{b}^{6}\alpha^{3}}
{3\pi m_{b}^{2}} \, |R_{nS}(0)|^{2} \left(1-\frac{12.6\alpha_{s}}{\pi}\right).
\end{split}
\end{equation}

The authors of Ref.~\cite{Ackleh:1991dy} give general expressions for singlet 
quarkonium decays into two gluons or two photons:
\begin{equation}
\begin{split}
\Gamma(n^{1}S_{0}\to 2g) &= \frac{2\alpha_{s}^{2}}{3m_{b}^{2}} \, 
|R_{nS}(0)|^{2} \left( 1+\frac{4.4\alpha_{s}}{\pi}\right), \\
\Gamma(n^{1}S_{0}\to 2\gamma) &= 
\frac{3e_{b}^{4}\alpha^{2}}{m_{b}^{2}} \, |R_{nS}(0)|^{2} \left( 
1-\frac{3.4\alpha_{s}}{\pi}\right). \\
\end{split}
\end{equation}

The annihilation decay widths of the $P$-wave bottomonium states depend on the 
derivative of the radial wave function at the origin. The relevant expressions 
have been summarize in Refs.~\cite{Kwong:1987ak,Kwong:1988ae} and are given here 
for completeness:
\begin{equation}
\begin{split}
\Gamma(n^{3}P_{0}\to 2\gamma) &= \frac{27e_{b}^{4}\alpha^{2}}{m_{b}^{4}} 
\, |R'_{nP}(0)|^{2} \left(1+\frac{0.2\alpha_{s}}{\pi} \right), \\
\Gamma(n^{3}P_{2}\to 2\gamma) &= \frac{36e_{b}^{4}\alpha^{2}}{5m_{b}^{4}} 
\, |R'_{nP}(0)|^{2} \left(1-\frac{16\alpha_{s}}{3\pi}\right), \\
\end{split}
\end{equation}
for their annihilation into photons, and
\begin{equation}
\begin{split}
\Gamma(n^{3}P_{0}\to 2g) &= \frac{6\alpha_{s}^{2}}{m_{b}^{4}} 
\, |R'_{nP}(0)|^{2}, \\
\Gamma(n^{3}P_{2}\to 2g) &=
\frac{8\alpha_{s}^{2}}{5m_{b}^{4}} \, |R'_{nP}(0)|^{2}, \\
\Gamma(n^{3}P_{1}\to q\bar{q}+g) &=
\frac{8n_{f}\alpha_{s}^{3}}{9\pi m_{b}^{4}} \, |R'_{nP}(0)|^{2} \,
\ln(m_{b}\left\langle r \right\rangle), \\
\Gamma(n^{1}P_{1}\to 3g) &=
\frac{20\alpha_{s}^{3}}{9\pi m_{b}^{4}} \, |R'_{nP}(0)|^{2} \,
\ln(m_{b}\left\langle r \right\rangle), \\
\end{split}
\end{equation}
for their annihilation into gluons and light quarks. We do not take into 
account the QCD corrections since these depend on each state and they are not 
known for the higher excited states. Moreover, one expects that these 
corrections are small as they concern to the bottomonium spectrum.

The decay rates for $\Upsilon(^{3}D_{J})\to 3g$ are dominated to leading order 
in logarithms by processes in which one of the three gluons is soft. (Two 
gluons cannot be emitted by a $^{3}D_{J}$ state since the charge-conjugation 
eigenvalue of a $^{3}D_{J}$ state is odd). The resulting expressions for the 
decay widths are~\cite{Belanger:1987cg}
\begin{equation}
\begin{split}
\Gamma(n^{1}D_{2}\to 2g) &= \frac{2\alpha_{s}^{2}}{3\pi m_{b}^{6}}
\, |R_{nD}''(0)|^{2}, \\
\Gamma(n^{3}D_{1}\to 3g) &=
\frac{760\alpha_{s}^{3}}{81\pi m_{b}^{6}} \, |R_{nD}''(0)|^{2} \,
\ln\left(4m_{b}\left\langle r \right\rangle\right), \\
\Gamma(n^{3}D_{2}\to 3g) &= 
\frac{10\alpha_{s}^{3}}{9\pi m_{b}^{6}} \, |R_{nD}''(0)|^{2} \,
\ln\left(4m_{b}\left\langle r \right\rangle\right), \\
\Gamma(n^{3}D_{3}\to 3g) &= 
\frac{40\alpha_{s}^{3}}{9\pi m_{b}^{6}} \, |R_{nD}''(0)|^{2} \,
\ln\left(4m_{b}\left\langle r \right\rangle\right).
\end{split}
\end{equation}

It is important to remark here that these formulas should be regarded as 
estimates of the partial widths for these annihilation processes rather than 
precise predictions. This is because considerable uncertainties arise in these 
expressions from the model-dependence of the wave functions and possible 
relativistic and QCD radiative corrections. 

Finally, an important progress has been done within pNRQCD in the computation 
of the inclusive decay widths into light hadrons, photons and lepton pairs for 
$S$- and $P$-wave heavy quarkonium states~\cite{Brambilla:2001xy, 
Brambilla:2002nu} (see also Ref.~\cite{Vairo:2003gh} for a review). These 
expressions need of $6$ non-perturbative universal parameters plus the 
knowledge of the heavy quarkonium wave functions (and their derivatives) at the 
origin. All these unknown terms should be fixed by experiment or be computed in 
lattice QCD in order to avoid model dependences.

%%%%%%%%%%%%%%%%%%%%%%%%%%%%%%%%%%%%%%%%%%%%%%%%%%%%%%%%%%%%%%%%%%%%%%%%%%%%%%%%

\subsection{Open-flavor meson strong decays}
\label{app:decas3P0model}

Meson strong decay is a complex nonperturbative process that has not yet been
described from first principles of QCD. Several phenomenological models have
been developed to deal with this topic (see, for instance,
Ref.~\cite{Segovia:2013kg} for a recent development). The most popular is the
$^{3}P_{0}$ model~\cite{Micu:1968mk,LeYaouanc:1972ae,LeYaouanc:1973xz} which
assumes that a quark-antiquark pair is created with vacuum quantum numbers,
$J^{PC}=0^{++}$.

An important characteristic of the $^{3}P_{0}$ model, apart from its simplicity, 
is that it provides the gross features of various transitions with only one 
parameter, the strength $\gamma$ of the decay interaction. Some attempts have 
been done to find possible dependences of the vertex parameter $\gamma$,
see~\cite{Ferretti:2013vua} and references therein. In 
Ref.~\cite{Segovia:2012cd} we performed a global fit to the decay widths of the
mesons which belong to charmed, charmed-strange, hidden charm and hidden bottom
sectors and elucidated the dependence on the mass scale of the $^{3}P_{0}$ free
parameter $\gamma$. Further details about the global fit can be found in
Ref.~\cite{Segovia:2012cd}. The running of the strength $\gamma$ of the
$^{3}P_{0}$ decay model is given by
\begin{equation}
\gamma(\mu) = \frac{\gamma_{0}}{\log\left(\frac{\mu}{\mu_{0}}\right)},
\label{eq:fitgamma}
\end{equation}
where $\mu$ is the reduced mass of the quark-antiquark in the decaying meson
and, $\gamma_{0}=0.81\pm0.02$ and $\mu_{0}=(49.84\pm2.58)\,{\rm MeV}$ are
parameters determined by the global fit.

We get a quite reasonable global description of the total decay widths in all
meson sectors, from light to heavy. All the wave functions for the mesons
involved in the open-flavor strong decays are the solutions of the
Schr\"odinger equation with the potential model described above and using the
Gaussian Expansion Method~\cite{Hiyama:2003cu}. We use when possible
experimental masses of the mesons involved in the open-flavor strong decays.
This is a standard procedure within the quark model approach and allows one to
ensure correct phase-space of the transition. Details of the resulting matrix
elements for different cases are given in Ref.~\cite{SegoviaThesis}, here we
proceed to explain briefly the main ingredients in which the model is based.

\subsubsection{Transition operator}

The interaction Hamiltonian involving Dirac quark fields that describes the
production process is given by
\begin{equation}
H_{I}=\sqrt{3}\,g_{s}\int d^{3}x \, \bar{\psi}(\vec{x})\psi(\vec{x}),
\label{eq:IH3P0}
\end{equation}
where we have introduced for convenience the numerical factor $\sqrt{3}$, which
will be canceled with the color factor.

If we write the Dirac fields in second quantization and keep only the
contribution of the interaction Hamiltonian which creates a $(\mu\nu)$
quark-antiquark pair, we arrive, after a nonrelativistic reduction, to the
following expression for the transition operator
\begin{equation}
\begin{split}
T =& -\sqrt{3} \, \sum_{\mu,\nu}\int d^{3}\!p_{\mu}d^{3}\!p_{\nu}
\delta^{(3)}(\vec{p}_{\mu}+\vec{p}_{\nu})\frac{g_{s}}{2m_{\mu}}\sqrt{2^{5}\pi}
\,\times \\
&
\times \left[\mathcal{Y}_{1}\left(\frac{\vec{p}_{\mu}-\vec{p}_{\nu}}{2}
\right)\otimes\left(\frac{1}{2}\frac{1}{2}\right)1\right]_{0}a^{\dagger}_{\mu}
(\vec{p}_{\mu})b^{\dagger}_{\nu}(\vec{p}_{\nu}),
\label{eq:Otransition2}
\end{split}
\end{equation}
where $\mu$ $(\nu)$ are the spin, flavor and color quantum numbers of the
created quark (antiquark). The spin of the quark and antiquark is coupled to
one. The ${\cal Y}_{lm}(\vec{p}\,)=p^{l}Y_{lm}(\hat{p})$ is the solid harmonic
defined in function of the spherical harmonic.

As in Ref.~\cite{Ackleh:1996yt}, we fix the relation of $g_{s}$ with the
dimensionless constant giving the strength of the quark-antiquark pair creation
from the vacuum as $\gamma=g_{s}/2m$, being $m$ the mass of the created quark
(antiquark). In this convention, values of the scale-dependent strength $\gamma$
in the different quark sectors following Eq.~(\ref{eq:fitgamma}) can be found in
Ref.~\cite{Segovia:2012cd}.

\begin{figure}[!t]
\begin{center}
\epsfig{figure=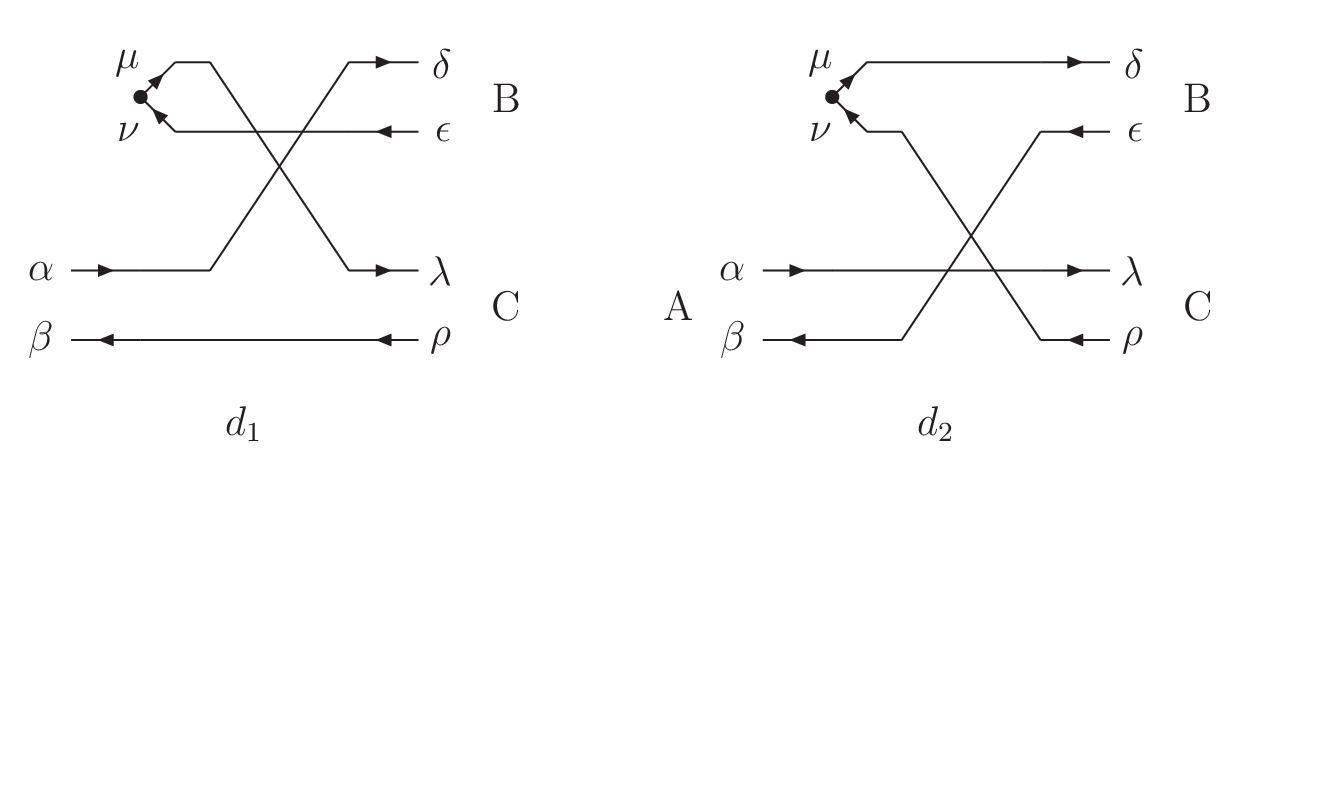,height=3.75cm,width=8.5cm}
\caption{\label{fig:3P0diagrams} Diagrams that can contribute to the decay
width through the $^{3}P_{0}$ model.}
\end{center}
\end{figure}

\subsubsection{Transition amplitude}

We are interested on the transition amplitude for the reaction
$(\alpha\beta)_{A} \to (\delta\epsilon)_{B} + (\lambda\rho)_{C}$. The meson $A$
is formed by a quark $\alpha$ and antiquark $\beta$. At some point it is created
a $(\mu\nu)$ quark-antiquark pair. The created $(\mu\nu)$ pair together with the
$(\alpha\beta)$ pair in the original meson regroups in the two outgoing mesons
via a quark rearrangement process. These final mesons are meson $B$ which is
formed by the quark-antiquark pair $(\delta\epsilon)$ and meson $C$ with
$(\lambda\rho)$ quark-antiquark pair.

We work in the center-of-mass reference system of meson $A$, thus we have
$\vec{K}_{A}=\vec{K}_{0}=0$ with $\vec{K}_{A}$ and $\vec{K}_{0}$ the total
momentum of meson $A$ and of the system $BC$ with respect to a given reference
system. We can factorize the matrix element as follows
\begin{equation}
\left\langle BC|T|A\right\rangle = \delta^{(3)}(\vec{K}_{0})
\mathcal{M}_{A\rightarrow BC}.
\end{equation}

The initial state in second quantization is
\begin{equation}
\left.|A\right\rangle
=\int d^{3}p_{\alpha}d^{3}p_{\beta}\delta^{(3)}(\vec{K}_{A}-\vec{P}_{A})
\phi_{A}(\vec{p}_{A})a_{\alpha}^{\dagger}(\vec{p}_{\alpha})b_{\beta}^{\dagger}
(\vec{p}_{\beta})\left.|0\right\rangle,
\label{eq:Istate}
\end{equation}
where $\alpha$ $(\beta)$ are the spin, flavor and color quantum numbers of
the quark (antiquark). The wave function $\phi_{A}(\vec{p}_{A})$ denotes a meson
$A$ in a color singlet with an isospin $I_{A}$ with projection $M_{I_{A}}$, a
total angular momentum $J_{A}$ with projection $M_{A}$, $J_{A}$ is the coupling
of angular momentum $L_{A}$ and spin $S_{A}$. The $\vec{p}_{\alpha}$ and
$\vec{p}_{\beta}$ are the momentum of quark and antiquark, respectively. The
$\vec{P}_{A}$ and $\vec{p}_{A}$ are the total and relative momentum of the
$(\alpha\beta)$ quark-antiquark pair within the meson $A$. The final state is
more complicated than the initial one because it is a two-meson state. It can be
written as
\begin{widetext}
\begin{equation}
\begin{split}
|BC\!\!\left.\right\rangle =& \frac{1}{\sqrt{1+\delta_{BC}}}\int d^{3}K_{B}
d^{3}K_{C}\sum_{m,M_{BC}}\left\langle\right.
\!\!J_{BC}M_{BC}lm|J_{T}M_{T}\!\!\left.\right\rangle\delta^{(3)}(\vec{K}-\vec{K_
{0}})\delta(k-k_{0}) \\ 
& 
\frac{Y_{lm}(\hat{k})}{k}\sum_{M_{B},M_{C},M_{I_{B}},M_{I_{C}}}\left\langle
J_{B}M_{B}J_{C}M_{C}|J_{BC}M_{BC}\right\rangle\left\langle
I_{B}M_{I_{B}}I_{C}M_{I_{C}}|I_{A}M_{I_{A}} \right\rangle \\ 
& 
\int d^{3}p_{\delta}d^{3}p_{\epsilon}d^{3}p_{\lambda}d^{3}p_{\rho}
\delta^{(3)}(\vec{K}_{B}-\vec{P}_{B})\delta^{(3)}(\vec{K}_{C}-\vec{P}_{C}) \\ 
&
\phi_{B}(\vec{p}_{B})\phi_{C}(\vec{p}_{C})a_{\delta}^{\dagger}(\vec{p}_{\delta}
)b_{\epsilon}^{\dagger}(\vec{p}_{\epsilon})a_{\lambda}^{\dagger}(\vec{p}_{
\lambda})b_{\rho}^{\dagger}(\vec{p}_{\rho})\left.|0\right\rangle,
\label{eq:Fstate}
\end{split}
\end{equation}
\end{widetext}
where we have followed the notation of meson $A$ for the mesons $B$ and $C$. We
assume that the final state of mesons $B$ and $C$ is a spherical wave with
angular momentum $l$. The relative and total momentum of mesons $B$ and $C$ are
$\vec{k}_{0}$ and $\vec{K}_{0}$. The total spin $J_{BC}$ is obtained coupling
the total angular momentum of mesons $B$ and $C$, and $J_{T}$ is the coupling of
$J_{BC}$ and $l$.

The $^{3}P_{0}$ model takes into account only diagrams in which the $(\mu\nu)$
quark-antiquark pair separates into different final mesons. This was originally
motivated by the experiment and it is known as the Okubo-Zweig-Iizuka
(OZI)-rule~\cite{okubo1963,zweigcern2,iizuka1966systematics} which tells us
that the disconnected diagrams are more suppressed than the connected ones. The
diagrams that can contribute to the decay width through the $^{3}P_{0}$ model
are shown in Fig.~\ref{fig:3P0diagrams}.

\subsubsection{Decay width}

The total width is the sum over the partial widths characterized by the quantum 
numbers $J_{BC}$ and $l$
\begin{equation}
\Gamma_{A\rightarrow BC}=\sum_{J_{BC},l}\Gamma_{A\rightarrow BC}(J_{BC},l),
\end{equation}
where
\begin{equation}
\Gamma_{A\rightarrow BC}(J_{BC},l)=2\pi\int
dk_{0}\delta(E_{A}-E_{BC})|\mathcal{M}_{A\rightarrow BC}(k_{0})|^{2}.
\label{eq:gammaDelta}
\end{equation}
We use relativistic phase space, so
\begin{equation}
\begin{split}
\Gamma_{A\rightarrow 
BC}(J_{BC},l)=2\pi\frac{E_{B}(k_{0})E_{C}(k_{0})}{m_{A}k_{0}}|\mathcal{M}_{
A\rightarrow BC}(k_{0})|^{2},
\end{split}
\end{equation}
where
\begin{equation}
k_{0}=\frac{\sqrt{[m_{A}^{2}-(m_{B}-m_{C})^{2}][m_{A}^{2}-(m_{B}+m_{C})^{2}]}}{
2m_{A}},
\end{equation}
is the on-shell relative momentum of mesons $B$ and $C$.

%%%%%%%%%%%%%%%%%%%%%%%%%%%%%%%%%%%%%%%%%%%%%%%%%%%%%%%%%%%%%%%%%%%%%%%%%%%%%%%%

\subsection{Hadronic decays}

The general way of referring to an hadronic transition 
is~\cite{Brambilla:2010cs} 
\begin{equation}
\Phi_{I} \to \Phi_{F}+h,
\end{equation}
where $\Phi_{I}$ and $\Phi_{F}$ stand, respectively, for the initial and final 
states of heavy quarkonium. The light hadron(s), $h$, are converted from 
emitted gluons and are kinematically dominated by single particle ($\pi^{0}$, 
$\eta$, $\omega$, $\ldots$) or two particle ($2\pi$, $2K$, $\ldots$) states. 

Since the energy difference between the initial and final quarkonium states is 
usually small, the emitted gluons are rather soft. In 
Ref.~\cite{Gottfried:1977gp}, Gottfried pointed out that this gluon radiation 
can be treated in a multipole expansion since the wavelengths of the emitted 
gluons are large compared with the size of the heavy quarkonium states. After 
the expansion of the gluon field, the Hamiltonian of the system can be 
decomposed as follows
\begin{equation}
{\cal H}^{\rm eff}_{\rm QCD} = {\cal H}^{(0)}_{\rm QCD} + {\cal H}^{(1)}_{\rm
QCD} + {\cal H}^{(2)}_{\rm QCD},
\label{eq:Hqcd}
\end{equation}
with ${\cal H}^{(0)}_{\rm QCD}$ the sum of the kinetic and potential energies of
the heavy quarkonium, and ${\cal H}^{(1)}_{\rm QCD}$ and ${\cal H}^{(2)}_{\rm
QCD}$ defined by 
\begin{equation}
\begin{split}
{\cal H}^{(1)}_{\rm QCD} &= Q_{a} A^{a}_{0}(x,t), \\
{\cal H}^{(2)}_{\rm QCD} &=-d_{a} E^{a}(x,t) - m_{a} B^{a}(x,t),
\label{eq:Hqcd2}
\end{split}
\end{equation}
in which $Q_{a}$, $d_{a}$ and $m_{a}$ are the color charge, the color electric
dipole moment and the color magnetic dipole moment, respectively. As the
$Q\bar{Q}$ pair is a color singlet, there is no contribution of the ${\cal
H}^{(1)}_{\rm QCD}$ and only $E_{l}$ and $B_{m}$ transitions can take place.

The multipole expansion within QCD (QCDME) has been studied by many 
authors~\cite{Gottfried:1977gp,Bhanot:1979af,Peskin:1979va,Bhanot:1979vb,
Voloshin:1978hc,Voloshin:1980zf}, but Tung-Mow Yan was the first one to present 
a gauge-invariant formulation in Refs.~\cite{Yan:1980uh,Kuang:1981se}. We will 
follow the updated review~\cite{Kuang:2006me} and references therein to 
calculate the hadronic transitions in which we are interested. A brief 
description of the formulae can be found below.

\subsubsection{Spin-nonflip $\pi\pi$ and $\eta$ transitions}
\label{subsubsec:Spinnonflip}

The spin-nonflip $\pi\pi$ decays in heavy quarkonia are dominated by double
electric-dipole transitions (E1E1). Therefore, the transition amplitude can be 
written as follows~\cite{Kuang:2006me}
\begin{equation}
{\cal M}_{E1E1}=i\frac{g_{E}^{2}}{6} \left\langle\right.\!\! \Phi_{F}h \,
|\vec{x}\cdot\vec{E} \, \frac{1}{E_{I}-H^{(0)}_{QCD}-iD_{0}} \,
\vec{x}\cdot\vec{E}| \, \Phi_{I} \!\! \left.\right\rangle,
\label{eq:E1E1}
\end{equation}
where $\vec{x}$ is the separation between $Q$ and $\bar{Q}$, and
$(D_0)_{bc}\equiv\delta_{bc}\partial_{0}-g_{s}f_{abc}A^{a}_{0}$.

Inserting a complete set of intermediate states the transition
amplitude~(\ref{eq:E1E1}) becomes
\begin{equation}
{\cal M}_{E1E1}=i\frac{g_{E}^{2}}{6} \sum_{KL}
\frac{\left\langle\right.\!\! \Phi_{F}|x_k|KL \!\!\left.\right\rangle
\left\langle\right.\!\! KL|x_l|\Phi_I \!\!\left.\right\rangle}{E_I-E_{KL}}
\left\langle\right.\!\! \pi\pi|E^{a}_{k} E^{a}_{l}|0 \!\!\left.\right\rangle,
\label{eq:factorizedE1E1}
\end{equation}
where $E_{KL}$ is the energy eigenvalue of the intermediate state $|KL\rangle$
with the principal quantum number $K$ and the orbital angular momentum $L$.

The intermediate states in the hadronic transition are those produced after the
emission of the first gluon and before the emission of the second one. They are
color singlet states with a gluon and a color-octet $Q\bar{Q}$ pair and thus 
these states are the so-called hybrid mesons. It is difficult to calculate these
hybrid states from first principles of QCD and thus we take a reasonable 
model which will be explained below.

One can see in Eq.~(\ref{eq:factorizedE1E1}) that the transition amplitude 
splits into two factors. The first one concerns to the wave functions and 
energies of the initial and final quarkonium states as well as those of the 
intermediate hybrid mesons. All these quantities can be calculated using 
suitable quark models. The second one describes the conversion of the emitted 
gluons into light hadrons. As the momenta involved are very low this matrix 
element cannot be calculated using perturbative QCD and one needs to resort to 
a phenomenological approach based on soft-pion techniques~\cite{Brown:1975dz}. 
In the center-of-mass frame, the two pion momenta $q_{1}$ and $q_{2}$ are the 
only independent variables describing this matrix element which, in the 
nonrelativistic limit, can be parametrized
as~\cite{Brown:1975dz,Yan:1980uh,Kuang:1981se,Kuang:2006me}
\begin{equation}
\begin{split}
& \frac{g_{E}^{2}}{6} \left\langle\right.\!\!
\pi_{\alpha}(q_{1})\pi_{\beta}(q_{2})|E^{a}_{k}E^{a}_{l}|0
\!\!\left.\right\rangle =
\frac{\delta_{\alpha\beta}}{\sqrt{(2\omega_{1})(2\omega_ {2})}} \,\times \\
&
\times
\left[C_{1}\delta_{kl}q^{\mu}_{1}q_{2\mu} + C_{2}\left(q_{1k}q_{2l}+q_{1l}q_{2k}
-\frac{2}{3}\delta_{kl}\vec{q}_{1}\cdot\vec{q}_{2}\right)\right],
\label{HofE1E1}
\end{split}
\end{equation}
where $C_{1}$ and $C_{2}$ are two unknown constants. The $C_{1}$ term is
isotropic, while the $C_{2}$ term has a $L=2$ angular dependence. Thus, $C_{1}$
only contributes to the $S$-wave into $S$-wave transitions and we fix it 
through the $\Upsilon(2S)\to \Upsilon(1S)\pi\pi$ reaction. The $C_{2}$ 
parameter is fixed through the decay $\Upsilon_{2}(1D)\to \Upsilon(1S)\pi\pi$.

Finally, the transition rate is given by~\cite{Kuang:1981se}
\begin{widetext}
\begin{equation}
\begin{split}
\Gamma\left(\Phi_{I}(^{2s+1}{l_{I}}_{J_{I}}) \to
\Phi_{F}(^{2s+1}{l_{F}}_{J_{F}}) + \pi\pi\right) = &
\delta_{l_{I}l_{F}}\delta_{J_{I}J_{F}} (G|C_{1}|^{2}-\frac{2}{3}H|C_{2}|^{2}
)\left|\sum_{L}(2L+1) \left(\begin{matrix} l_{I} & 1 & L \\ 0 & 0 & 0
\end{matrix}\right) \left(\begin{matrix} L & 1 & l_{I} \\ 0 & 0 & 0
\end{matrix}\right) f_{IF}^{L11}\right|^{2} \\
&
+(2l_{I}+1)(2l_{F}+1)(2J_{F}+1) \sum_{k} (2k+1) (1+(-1)^{k})
\left\lbrace\begin{matrix} s & l_{F} & J_{F} \\ k & J_{I} & l_{I}
\end{matrix}\right\rbrace^{2} H |C_{2}|^{2} \times \\
&
\times\left|\sum_{L} (2L+1) \left(\begin{matrix} l_{F} & 1 & L \\ 0 & 0 & 0
\end{matrix}\right) \left(\begin{matrix} L & 1 & l_{I} \\ 0 & 0 & 0
\end{matrix}\right) \left\lbrace\begin{matrix} l_{I} & L & 1 \\ 1 & k & l_{F}
\end{matrix}\right\rbrace f_{IF}^{L11} \right|^{2},
\label{eq:gamapipi}
\end{split}
\end{equation}
with
\begin{equation}
f_{IF}^{LP_{I}P_{F}} = \sum_{K} \frac{1}{M_{I}-M_{KL}} \left[\int dr\,
r^{2+P_{F}} R_{F}(r)R_{KL}(r)\right] \left[\int dr' r'^{2+P_{I}} R_{KL}(r')
R_{I}(r')\right],
\label{eq:fifl}
\end{equation}
\end{widetext}
where $R_{I}(r)$ and $R_{F}(r)$ are, respectively, the radial wave functions of 
the initial and final states. $R_{KL}(r)$ is the radial wave function of the 
intermediate vibrational states $\left|KL\right\rangle$. The mass of the 
decaying meson is $M_{I}$, whereas the ones corresponding to the hybrid states 
are $M_{KL}$. The quantities $G$ and $H$ are the phase-space integrals
\begin{equation}
\begin{split}
G=&\frac{3}{4}\frac{M_{F}}{M_{I}}\pi^{3}\int
dM_{\pi\pi}^{2}\,K\,\left(1-\frac{4m_{\pi}^{2}}{M_{\pi\pi}^{2}}\right)^{1/2}(M_{
\pi\pi}^{2}-2m_{\pi}^{2})^{2}, \\
H=&\frac{1}{20}\frac{M_{F}}{M_{I}}\pi^{3}\int
dM_{\pi\pi}^{2}\,K\,\left(1-\frac{4m_{\pi}^{2}}{M_{\pi\pi}^{2}}\right)^{1/2}
\times \\
&
\times\left[(M_{\pi\pi}^{2}-4m_{\pi}^{2})^{2}\left(1+\frac{2}{3}\frac{K^{2}}{M_{
\pi\pi}^{2}}\right)\right. \\
&
\left.\quad\,\, +\frac{8K^{4}}{15M_{\pi\pi}^{4}}(M_{\pi\pi}^{4}+2m_{\pi}^{2}
M_{\pi\pi}^{2}+6m_{\pi}^{4})\right],
\end{split}
\end{equation}
with $K$ given by
\begin{equation}
K = \frac{\sqrt{\left[(M_{I}+M_{F})^{2}-M_{\pi\pi}^{2}\right]
\left[(M_{I}-M_{F})^{2}-M_{ \pi\pi}^{2}\right]}}{2M_{I}}.
\end{equation}

The leading multipoles of spin-nonflip $\eta$ transitions between spin-triplet
$S$-wave states are M1M1 and E1M2. Therefore, the matrix element is given
schematically by
\begin{equation}
{\cal M}(^{3}S_{1}\to\,^{3}\!S_{1} + \eta) = {\cal M}_{M1M1} +
{\cal M}_{E1M2}.
\end{equation}
After some algebra and assuming that ${\cal M}_{M1M1}=0$ (see
Ref.~\cite{Kuang:1981se} for details), the decay rate can be written as
\begin{equation}
\Gamma(\Phi_{I}(^{3}S_{1}) \to \Phi_{F}(^{3}S_{1}) + \eta) =
\frac{8\pi^{2}}{27} \frac{M_{f}C_{3}^{2}}{M_{i}m_{Q}^{2}} |f_{IF}^{111}|^{2}
|\vec{q}|^{3},
\end{equation}
where $\vec{q}$ is the momentum of $\eta$, $C_{3}$ is a new parameter which
should be fixed through the $\Upsilon(2S)\to\Upsilon(1S)\eta$ reaction. The 
function $f_{IF}^{111}$ is defined in Eq.~(\ref{eq:fifl}).

\subsubsection{Spin-flip $\pi\pi$ and $\eta$ transitions}
\label{subsubsec:spinflip}

The spin-flip $\pi\pi$ and $\eta$ transitions between heavy quarkonia are 
induced by an E1M1 multipole amplitude. Within the hadronization approach 
presented above, the description of this kind of decays implies the
introduction of another phenomenological constant which should be fixed by 
experiment. Therefore, as one can deduce, the decay model for hadronic 
transitions begins to loose its predictive power.

In order to avoid this undesirable feature, the term which describes the 
conversion of the emitted gluons into light hadrons can be computed assuming 
a duality argument between the physical light hadron final state and the 
associated two-gluon final 
state~\cite{Kuang:1981se}:
\begin{equation}
\begin{split}
\Gamma(\Phi_{I}\to \Phi_{F} + \pi\pi) &\sim \Gamma(\Phi_{I}\to \Phi_{F}gg), \\
\Gamma(\Phi_{I}\to \Phi_{F} + \eta) &\sim \Gamma(\Phi_{I}\to 
\Phi_{F}(gg)_{0^{-}}), \\
\end{split}
\end{equation}
where in the second line the two gluons are projected into a $J^{P}=0^{-}$ state
to simulate the $\eta$ meson. The advantage of this approach is that we have 
now only two free parameters, $g_{E}$ and $g_{M}$, in order to fix the 
spin-nonflip and spin-flip $\pi\pi$ and $\eta$ hadronic transitions.

Explicit expressions within this new approach of the decay rates for the 
spin-nonflip $\pi\pi$ and $\eta$ transitions can be found in 
Refs.~\cite{Kuang:1981se,Kuang:2006me}. The 
decay rates for the spin-flip $\pi\pi$ and $\eta$ transitions are
\begin{equation}
\begin{split}
&
\Gamma(\Phi_{I}(^{3}{l_{I}}_{J_{I}})\to \Phi_{F}(^{1}{l_{F}}_{J_{F}}) +
\pi\pi) = \left(\frac{g_{E}g_{M}}{6m_{Q}}\right)^{2} \times \\
&
\hspace*{0.40cm} \times \frac{(M_{i}-M_{f})^{7}}{315\pi^{3}} (2l_{F}+1)
\left(\begin{matrix} l_{F} & 1 & l_{I} \\ 0 & 0 & 0 \end{matrix}\right)^{2}
|f_{IF}^{l_{F}10}+f_{IF}^{l_{I}01}|^{2}, \\[2ex]
&
\Gamma(\Phi_{I}(^{3}{S}_{J_{I}})\to \Phi_{F}(^{1}{P}_{J_{F}}) + \eta) =
\frac{g_{M}^{2}}{g_{E}^{2}} \frac{E_{F}}{M_{I}} |\vec{q}| \times \\
&
\hspace*{0.40cm} \times \frac{\pi}{1144m_{Q}^{2}} 
\left(\frac{4\pi}{\sqrt{6}}f_{\pi}m_{\eta}^{2}\right)^{2}
|f_{IF}^{110}+f_{IF}^{001}|^{2}.
\label{eq:spinflip}
\end{split}
\end{equation}
The decay rate of the spin-flip $\eta$ transition in Eq.~(\ref{eq:spinflip}) 
can be read from the decay rate of the the isospin violating hadronic 
transition~\cite{Kuang:2006me}
\begin{equation}
\begin{split}
&
\Gamma(\Phi_{I}(^{3}{S}_{J_{I}})\to \Phi_{F}(^{1}{P}_{J_{F}}) + \pi^{0}) =
\frac{g_{M}^{2}}{g_{E}^{2}} \frac{E_{F}}{M_{I}} |\vec{q}| \times \\
&
\hspace*{0.40cm} \times \frac{\pi}{1144m_{Q}^{2}} 
\left(\frac{4\pi}{\sqrt{2}}\frac{m_{d}-m_{u}}{m_{d}+m_{u}}f_{\pi}m_{\pi}^{2}
\right)^{2} |f_{IF}^{110}+f_{IF}^{001}|^{2},
\end{split}
\end{equation}
in which the factor $(m_{d}-m_{u})/(m_{d}+m_{u})$ reflects the violation of
isospin.

\subsubsection{A model for hybrid mesons}
\label{subsubsec:hybrids}

From the generic properties of QCD, we might expect to have states in which the
gluonic field itself is excited and carries $J^{PC}$ quantum numbers. A
bound-state is called glueball when any valence quark content is absent, the
addition of a constituent quark-antiquark pair to an excited gluonic field gives
rise to what is called a hybrid meson. The gluonic quantum numbers couple to
those of the $q\bar{q}$ pair. This coupling may give rise to the so-called
exotic $J^{PC}$ mesons, but also can produce hybrid mesons with natural quantum
numbers. We are interested on the last ones because they are involved in the
calculation of hadronic transitions within the QCDME approach.

Ab-initio QCD calculations of the hybrid (even conventional) excited 
bottomonium states are particularly difficult because the large mass of the 
$b$-quark. For instance, full lattice QCD results can be found only for the 
charmonium hybrids in Ref.~\cite{Liu:2012ze} and the first application of the 
effective field theory pNRQCD to the hybrid meson spectrum has been published 
very recently in Ref.~\cite{Berwein:2015vca}.

An extension of the quark model described above to include hybrid states has 
been presented in Ref.~\cite{Segovia:2014mca}. This extension is inspired on 
the Buchmuller-Tye quark-confining string (QCS) model~\cite{Tye:1975fz, 
Giles:1977mp, Buchmuller:1979gy}. The QCS model is defined by a relativistic-, 
gauge- and reparametrization-invariant action describing quarks interacting with 
color $SU(3)$ gauge fields in a two dimensional world sheet. It is assumed that 
the meson is composed of a quark and antiquark linked by an appropriate color 
electric flux line (the string).

The string can carry energy-momentum only in the region between the quark and
the antiquark. The string and the quark-antiquark pair can rotate as a unit and
also vibrate. Ignoring its vibrational motion, the equation which describes the
dynamics of the quark-antiquark pair linked by the string should be the usual
Schr\"odinger equation with a confinement potential. Gluon excitation effects
are described by the vibration of the string. These vibrational modes provide
new states beyond the naive meson picture.

A complete description of the model can be found in
Refs.~\cite{Tye:1975fz,Giles:1977mp,Buchmuller:1979gy}. We will give here only a
brief description of it. The dynamics of the string is defined by the action
\begin{equation}
\begin{split}
S &= \int^{\infty}_{-\infty} \, d^2u \, \sqrt{-g} \, \times \\
&
\times \bigg\{ \sum_j\bar{\psi}_j \left[\gamma_\mu
\tau^{\alpha\mu}\left(\frac{i}{2}\bar{\partial_{\alpha}}
-eB_{a\alpha}T^{\alpha}\right)-M_j\right] \psi_j \\
&
\hspace{0.50cm} -\frac{1}{4}F_{a\alpha\beta}F^{\alpha\beta}_a \bigg\},
\end{split}
\end{equation}
where $\psi_j(u)$ is a four-component Dirac field, $d^2u\sqrt{-g}$ is the
invariant volume element, $T^{a}=\lambda^{a}/2$ are the eight matrix generators 
of $SU(3)$ color and $B_{a\alpha}$ are the color gauge fields. From this 
action, in the nonrelativistic limit, one obtains the effective 
Hamiltonian~\cite{Giles:1977mp} composed of three terms (the quark, the string 
and the Coulomb):
\begin{equation}
\begin{split}
{\cal H} &= {\cal H}_q + {\cal H}_s + {\cal H}_c \\
&
=\int d\sigma\chi^+\left(M\beta-i\alpha_1 \partial_1\right)\chi \\
&
\hspace{0.50cm} +\int d\sigma \chi^+\beta \chi\frac{Mv^2}{2} \\
&
\hspace{0.50cm} +\frac{e^2}{2}\int d\sigma
d\sigma'\chi^+(\sigma)T^a\chi(\sigma)G(\sigma,\sigma')\chi^+(\sigma')
T^a\chi(\sigma'),
\end{split}
\end{equation}
which, in absence of vibrations and after quantization of the rotational modes,
leads to the following Schr\"odinger equation for the meson bound-states in the
center-of-mass frame
\begin{equation}
\left[2M-\frac{1}{M}\frac{\partial^2}{\partial
r^2}+kr-\frac{l(l+1)}{Mr^2}\right]
\psi(r) = E\psi(r).
\end{equation}

The coupled equations that describe the dynamics of the string and the quark
sectors are very non-linear so that there is no hope of solving them completely.
Then, to introduce the vibrational modes, we use the following approximation
scheme. First, we solve the string Hamiltonian (via de Bohr-Oppenheimer method)
to obtain the vibrational energies as a function of the interquark distance.
These are then inserted into the meson equation as an effective potential,
$V_{n}(r)$.

Assuming the quark mass to be very heavy so that the ends of the string are
fixed, the vibrational potential energy can be estimated using the
Bohr-Sommerfeld quantization to be~\cite{Giles:1977mp}
\begin{equation}
\begin{split}
V_{n}(r) = \sigma r\left\lbrace 1 + \frac{2n\pi}{\sigma
\left[(r-2d)^{2}+4d^{2}\right]} \right\rbrace^{1/2},
\end{split}
\end{equation}
where $d$ is the correction due to the finite quark mass 
\begin{equation}
d(m_{Q},r,\sigma,n)=\frac{\sigma r^{2}\alpha_{n}}{4(2m_{Q} + \sigma
r\alpha_{n})},
\end{equation}
being $\alpha_{n}$ a parameter related with the shape of the vibrating
string~\cite{Giles:1977mp}, and can take the values $1\leq\alpha_{n}^{2}\leq2$.
For $n=0$, $V_{n}(r)$ reduces to the naive $Q\bar{Q}$ one. 

In our quark model, the central part of the confining potential has the
following form
\begin{equation}
V_{\rm CON}^{\rm C}(r) = \frac{16}{3} [a_{c}(1-e^{-\mu_{c}r})-\Delta],
\end{equation}
and can be written as
\begin{equation}
V_{\rm CON}^{\rm C}(r) = \sigma(r)r + \mbox{cte,}
\end{equation}
where
\begin{equation}
\begin{split}
\sigma(r) &= \frac{16}{3}\,a_{c}\,\left(\frac{1-e^{-\mu_{c}r}}{r}\right), \\
\mbox{cte} &= -\frac{16}{3}\,\Delta.
\end{split}
\end{equation}
This means that our effective string tension, $\sigma(r)$, is not a constant but
depends on the interquark distance, $r$. In fact, it decreases with respect to
$r$ until it reaches the string breaking region.

Following the ideas of Ref.~\cite{Buchmuller:1979gy}, the potential for hybrid
mesons derived from our constituent quark model has the following
expression~\cite{Segovia:2014mca}
\begin{equation}
V_{\rm hyb}(r)=V_{\rm OGE}^{\rm C}(r) + V_{\rm CON}^{\rm C}(r) +
\left[V_{n}(r) - \sigma(r)r\right],
\label{eq:pothyb}
\end{equation}
where we have not taken into account the spin-dependent terms. $V_{\rm OGE}^{\rm
C}(r) + V_{\rm CON}^{\rm C}(r)$ is the naive quark-antiquark potential and
$V_{n}(r)$ is the vibrational one. We must subtract the term $\sigma(r)r$
because it appears twice, one in $V_{\rm CON}^{\rm C}(r)$ and the other one in
$V_{n}(r)$. This potential does not include new model parameters and depends 
only on those coming from the original quark model. In such sense, the 
calculation of the hybrid states is parameter-free. More explicitly, our
different contributions are
\begin{equation}
\begin{split}
V_{\rm OGE}^{\rm C}(r) &= -\frac{4\alpha_{s}}{3r}, \\
V_{\rm CON}^{\rm C}(r) &= \frac{16}{3} [a_{c}(1-e^{-\mu_{c}r})-\Delta], \\
V_{n}(r) &= \sigma(r)r \left\lbrace 1 + \frac{2n\pi}{\sigma(r)
\left[(r-2d)^{2}+4d^{2}\right]} \right\rbrace^{1/2},
\end{split}
\end{equation}
where
\begin{equation}
d(m_{Q},r,\sigma,n) =
\frac{\sigma(r)r^{2}\alpha_{n}}{4(2m_{Q}+\sigma(r)r\alpha_{n})}.
\end{equation}

An important feature of our hybrid model is that, just like the naive quark 
model, the hybrid potential has a threshold from which no more states can be 
found and so we have a finite number of hybrid states in the spectrum. Hybrid 
meson masses calculated in the bottomonium sector using our model are shown in 
Table~\ref{tab:hybridsbb}.

\begin{table}[!t]
\begin{center}
 \begin{tabular}{cccc}
\hline
\hline
\tstrut
K & $L=0$ & $L=1$ & $L=2$ \\
\hline
\tstrut
$1$  & $10571$ & $10785$ & $10921$ \\
$2$  & $10857$ & $10999$ & $11108$ \\
$3$  & $11063$ & $11175$ & $11267$ \\
$4$  & $11232$ & $11325$ & $11402$ \\
$5$  & $11374$ & $11452$ & $11519$ \\
$6$  & $11496$ & $11562$ & $11619$ \\
$7$  & $11600$ & $11657$ & $11706$ \\
$8$  & $11690$ & $11738$ & $11780$ \\
$9$  & $11766$ & $11807$ & $11843$ \\
$10$ & $11831$ & $11866$ & $11895$ \\
$11$ & $11885$ & $11913$ & - \\
$12$ & $11927$ & - & - \\
\hline
\multicolumn{4}{c}{Threshold = 11943 MeV} \\
\hline
\hline
\end{tabular}
\caption{\label{tab:hybridsbb} Hybrid meson masses, in MeV, calculated in the
$b\bar{b}$ sector. The variation of the parameter $\alpha_{n}$ which range
between $1<\alpha_{n}<\sqrt{2}$ modifies the energy as much as $30\,{\rm MeV}$,
we have taken $\alpha_{n}=\sqrt{1.5}$.}
\end{center}
\end{table}

%%%%%%%%%%%%%%%%%%%%%%%%%%%%%%%%%%%%%%%%%%%%%%%%%%%%%%%%%%%%%%%%%%%%%%%%%%%%%%%%
%%%%%%%%%%%%%%%%%%%%%%%%%%%%%%%%%%%%%%%%%%%%%%%%%%%%%%%%%%%%%%%%%%%%%%%%%%%%%%%%

\section*{References}
\bibliographystyle{apsrev}
\bibliography{bottomonium}

\end{document}